\documentclass[twocolumn]{revtex4}
\usepackage{mathbbold}
\usepackage{amsfonts}
\usepackage{amsmath}
\usepackage{amssymb}
\usepackage{charter}
\usepackage{graphicx}

\input{tcilatex}

\begin{document}

\title{Enhancing quantum entanglement and quantum teleportation for two-mode
squeezed vacuum state by local quantum-optical catalysis }
\author{Xue--xiang Xu$^{\dag }$}
\affiliation{Department of Physics, Jiangxi Normal University, Nanchang 330022, China\\
$^{\dag }$Corresponding author: xuxuexiang@jxnu.edu.cn }

\begin{abstract}
I theoretically investigate how the entanglement properties of a two-mode
squeezed vacuum state (TMSVS) can be enhanced by operating quantum-optical
catalysis on each mode of the TMSVS. The quantum-optical catalysis is simply
mixing one photon at the beam splitter and post-select the beam-splitter
(BS) output based on detection of one photon, first proposed by Lvovsky and
Mlynek [Phys. Rev. Lett. \textbf{88},\ 250401 (2002)]. I find that there
exists some enhancement in the entanglement properties (namely, entanglement
entropy, second-order Einstein-Podolsky-Rosen correlation, and the fidelity
of quantum teleportation) in certain parameter ranges spanned by the low
transmissivities of the BSs and the small squeezing parameter of the input
TMSVS.

\textbf{PACS: }03.67.-a, 05.30.-d, 42.50,Dv, 03.65.Wj

\textbf{Keywords:} Degree of entanglement; EPR correlation; Teleportation
fidelity; Quantum-optical catalysis
\end{abstract}

\maketitle

\section{Introduction}

Entangled resources are useful in quantum information processing, such as
quantum teleportation \cite{1}, metrology \cite{2}, and communications \cite%
{3}. Two-mode squeezed vacuum state (TMSVS) is one of the most popular (if
not the most) tools for quantum-enhanced optical interferometers or
continuous variable (CV) quantum information processing as it is a Gaussian
and entangled state \cite{4,5,6}. However, theoretical investigations have
shown that Gaussian entangled resources have some restrictions \cite{7,8}.
For example, entanglement distillation from two Gaussian entangled states is
impossible by Gaussian local operations and classical communication \cite%
{9,10}. Therefore, it is desirable to seek for non-Gaussian entangled
resources and operations which can be more efficient in the quantum
information processing. In recent years, some entanglement criteria beyond
the Gaussian regime, including all orders of Einstein-Podolsky-Rosen (EPR)
correlations, have been proposed \cite{11,12}. Moreover, it has been shown
that non-Gaussian two-mode entangled states provide the benefits of
enhancing the entanglement \cite{13,14,15,16}.

Previously, the effect on the entanglement has been theoretically analyzed
based on the merit of concrete protocol, such as the degree of entanglement,
the EPR correlation, and the fidelity of teleportation \cite{13,14}. In
fact, it was shown that the performance of every protocol was improved,
implying that the entanglement of a non-Gaussian state must be enhanced \cite%
{12}. In recent years, many schemes of generating two-mode non-Gaussian
entangled states have been proposed. Among these schemes, performing
non-Gaussian operation on a two-mode Gaussian state is a possible approach
to generate non-Gaussian entangled resources \cite{17,18}. These typical
non-Gaussian operations include the elementary operations (i.e., photon
addition $a^{\dag },b^{\dag }$ and subtraction $a,b$) and their sequential
operations (e.g., $ab,a^{\dag }b^{\dag }$) \cite{19,20} as well as their
coherent superposition (e.g., $a^{\dag 2}+b^{\dag 2}$) \cite{21}. Recently,
in order to implement multiple photon addition and subtraction on both modes
of the TMSVS, Navarrete-Benlloch \textit{et al.} \cite{15} demonstrate that
the entanglement generally increases with the number of such operations. On
the other hand, one can generate non-Gaussian entangled resources by means
of a linear or nonlinear quantum-optical system \cite{22,23,24}. The systems
generally consist of beam splitting, phase shifting, squeezing,
displacement, and various detection.

About two decades ago, the concept of \textquotedblleft conditional
measurement\textquotedblright\ was proposed by Dakna \textit{et al. }\cite%
{25}. They generate a Schrodinger-cat-like state\ by using a simple
beam-splitter (BS) scheme for a conditional measurement. Following Dakna's
idea of conditional measurement, many schemes have been proposed to prepare
quantum states \cite{26,27,28}. Among these works, the typical proposal is
the quantum-optical catalysis, proposed by Lvovsky and Mlynek \cite{29}.
They generated a coherent superposition state $t\left\vert 0\right\rangle
+\alpha \left\vert 1\right\rangle $ by conditional measurement on a BS. This
state was generated in one of the BS output channels if a coherent state $%
\left\vert \alpha \right\rangle $ and a single-photon Fock state $\left\vert
1\right\rangle $ are present in two input ports and a single photon is
registered in the other BS output. They call this transformation as
\textquotedblleft quantum-optical catalysis\textquotedblright\ because the
single photon itself remains unaffected but facilitates the conversion of
the target ensemble. Recently, Bartley \textit{et al. }\cite{30} perform
quantum-optical catalysis to generate multiphoton nonclassical states, which
create a wide range of nonclassical phenomena. Since performing
quantum-optical catalysis on a single-mode Gaussian state can enhance
nonclassicality of the given state, one can ask whether it is possible to
enhance entanglement of a two-mode Gaussian state via quantum-optical
catalysis. This issue will be addressed here.

In this paper, I propose a scheme to generate a two-mode non-Gaussian
entangled state. This state is generated by operating quantum-optical
catalysis on each mode of a TMSVS. I investigate the entanglement properties
(the degree of entanglement and EPR correlation) and the quantum
teleportation fidelity for the state I produce. I show that when ideal
quantum-optical catalysis is used, the input Gaussian state can be
transformed into a non-Gaussian state with higher entanglement.

The paper is organized as follows. In Sec.II, I begin with the generation of
a non-Gaussian two-mode entangled state by operating quantum-optical
catalysis from a two-mode squeezed vacuum state (TMSVS) and derive its
normalization factor (i.e., success probability), which is important to
discussing quantum properties. In Sec.III, I investigate the entanglement
properties (degree of entanglement and EPR correlation) of the non-Gaussian
state\ and analyze the effect of the local quantum-optical catalysis. Then,
I consider the non-Gaussian entangled state as an entangled resource to
teleport a coherent state in Sec.IV. The main results are summarized in
Sec.V.

\section{Two-mode non-gaussian entangled state by local quantum-optical
catalysis}

In this section, I make a brief review of quantum-optical catalysis\ and
apply it to prepare a two-mode non-Gaussian quantum state. The theoretical
scheme is proposed and the success probability is derived.

\subsection{Theoretical scheme}

The basic idea on the quantum-optical catalysis was introduced in Ref.\cite%
{29}. The conceptual schematic is shown in Fig. 1. If an input state $\rho
_{in}$ and a single-photon Fock state $\left\vert 1\right\rangle $ are
present in the two input ports of the BS and a single photon $\left\vert
1\right\rangle $\ is registered in one BS output port, then a catalyzed
state $\rho _{c}$ can be generated in the other BS output channel. 
\begin{figure}[tbp]
\label{Fig1} \centering\includegraphics[width=0.8\columnwidth]{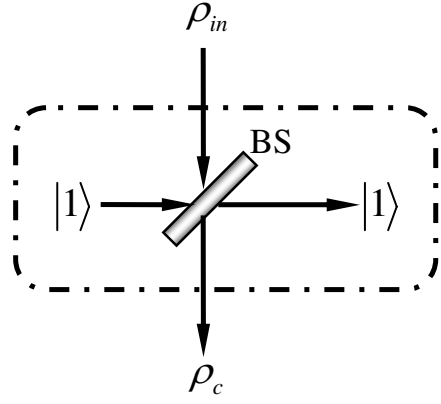}
\caption{Basic block of the quantum-optical catalysis. An input state $%
\protect\rho _{in}$ and a single-photon Fock state $\left\vert
1\right\rangle $ are present in the two input ports of the BS. Measurement
is conditioned on registering a single photon $\left\vert 1\right\rangle $\
by the single-photon detector. Here, the output state $\protect\rho _{c}$ is
called as the catalysis state of the input state $\protect\rho _{in}$\ and
this process\ is quantum-optical catalysis. The catalysis parameter is the
tunable transmissivity of the BS.}
\end{figure}

My scheme is depicted in Fig. 2. Theoretically, the input TMSVS $\left\vert
\psi _{0}\right\rangle _{ab}$ is obtained by applying the unitary operator $%
S_{2}\left( r\right) $ on the two-mode vacuum $\left\vert
0_{a},0_{b}\right\rangle $, i.e.%
\begin{eqnarray}
\left\vert \psi _{0}\right\rangle _{ab} &=&S_{2}\left( r\right) \left\vert
0_{a},0_{b}\right\rangle =\frac{1}{\cosh r}\exp \left( a^{\dag }b^{\dag
}\tanh r\right) \left\vert 0_{a},0_{b}\right\rangle   \notag \\
&=&\frac{1}{\cosh r}\sum_{n=0}^{\infty }\tanh ^{n}r\left\vert
n_{a},n_{b}\right\rangle ,  \label{1-1}
\end{eqnarray}%
where $S_{2}\left( r\right) =\exp \left[ r\left( a^{\dag }b^{\dag
}-ab\right) \right] $ is the two-mode squeezed operator and the values of $r$
determines the degree of squeezing. The larger $r$, the more the state is
squeezed. In particular, when $r=0$, $\left\vert \psi _{0}\right\rangle _{ab}
$ reduces to $\left\vert 0_{a},0_{b}\right\rangle $. Enlightened by the idea
of quantum-optical catalysis, I prepare a new state from the TMSVS $%
\left\vert \psi _{0}\right\rangle _{ab}$ by operating quantum-optical
catalysis\ on each mode. Then, the prepared state $\left\vert \psi
_{LQC}\right\rangle _{ab}$ is given by 
\begin{equation}
\left\vert \psi _{LQC}\right\rangle _{ab}=\frac{1}{\sqrt{p_{cd}}}%
\left\langle 1_{d}\right\vert \left\langle 1_{c}\right\vert
B_{2}B_{1}S_{2}\left( r\right) \left\vert 0_{a},0_{b}\right\rangle
\left\vert 1_{c}\right\rangle \left\vert 1_{d}\right\rangle ,  \label{1-2}
\end{equation}%
which will be called as \textquotedblleft local quantum catalyzed
TMSVS\textquotedblright\ (LQC-TMSVS). Here, $B_{1}$ and $B_{2}$\ correspond
to the respective unitary operators of the two tunable $BS_{1}$ and $BS_{2}$
with 
\begin{equation}
B_{1}=\exp \left[ \theta _{1}\left( a^{\dag }c-ac^{\dag }\right) \right] ,%
\text{ \ \ }B_{2}=\exp \left[ \theta _{2}\left( b^{\dag }d-bd^{\dag }\right) %
\right] .  \label{1-3}
\end{equation}%
in terms of the creation (annihilation) operator $a^{\dag }$ ($a$), $b^{\dag
}$ ($b$), $c^{\dag }$ ($c$), and $d^{\dag }$ ($d$)\ for modes $a,b,c,$ and $d
$. Using Eq.(\ref{1-3}), one obtains the following transformations:%
\begin{eqnarray}
B_{1}a^{\dag }B_{1}^{\dag } &=&a^{\dag }t_{1}-c^{\dag }r_{1},\text{ \ }%
B_{1}c^{\dag }B_{1}^{\dag }=a^{\dag }r_{1}+c^{\dag }t_{1},  \notag \\
B_{2}b^{\dag }B_{2}^{\dag } &=&b^{\dag }t_{2}-d^{\dag }r_{2},\text{ \ }%
B_{2}d^{\dag }B_{2}^{\dag }=b^{\dag }r_{2}+d^{\dag }t_{2}.  \label{1-4}
\end{eqnarray}%
where $t_{j}=\cos \theta _{j}$ and $r_{j}=\sin \theta _{j}\ (j=1,2)$\ are\
the transmission coefficient and the reflection coefficient\ of the beam
splitter $BS_{j}$, respectively. The normalization factor $p_{cd}$
represents the success probability heralded by the detection of a single
photon at the modes $c$ and $d$.

\begin{figure}[tbp]
\label{Fig2} \centering\includegraphics[width=0.9\columnwidth]{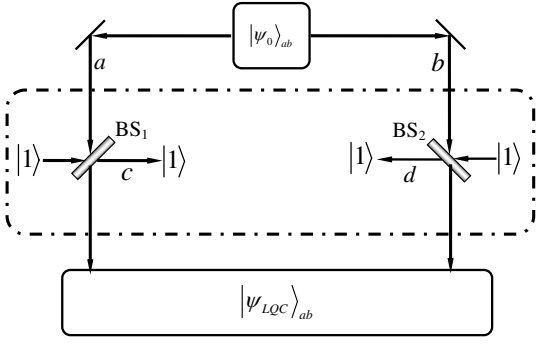}
\caption{Optical scheme to prepare a LQC-TMSVS by operating quantum-optical
catalysis on each mode of a TMSVS. The input state is the TMSVS $\left\vert 
\protect\psi _{0}\right\rangle _{ab}$ with the squeezing parameter $r$ and
the output state is the LQC-TMSVS $\left\vert \protect\psi %
_{LQC}\right\rangle _{ab}$ related with the input and catalysis parameter.
The catalysis parameters are determined by the transmissivities $T_{1}$ and $%
T_{2}$\ of the tunable $BS_{1}$ and $BS_{2}$, respectively. In contrast with
the TMSVS, the LQC-TMSVS has a wide range of entanglement properties.}
\end{figure}

Using above relation and some technique (see Appendix A), the LQC-TMSVS $%
\left\vert \psi _{LQC}\right\rangle _{ab}$ can be expressed explicitly as
follows%
\begin{equation}
\left\vert \psi _{LQC}\right\rangle _{ab}=\left( c_{0}\allowbreak
+c_{1}a^{\dag }b^{\dag }+c_{2}a^{\dag 2}\allowbreak b^{\dag 2}\right)
S_{2}\left( \lambda \right) \left\vert 0_{a},0_{b}\right\rangle ,
\label{1-5}
\end{equation}%
where a squeezing parameter $\lambda $ satisfying%
\begin{equation*}
\tanh \lambda =t_{1}t_{2}\tanh r,
\end{equation*}%
and $\ $%
\begin{eqnarray*}
c_{0} &=&\frac{t_{1}t_{2}\cosh \lambda }{\sqrt{p_{cd}}\cosh r}, \\
c_{1} &=&\frac{\left( r_{1}^{2}r_{2}^{2}-r_{1}^{2}\allowbreak
t_{2}^{2}-r_{2}^{2}t_{1}^{2}\right) \tanh r\cosh \lambda }{\sqrt{p_{cd}}%
\cosh r}, \\
c_{2} &=&\frac{r_{1}^{2}r_{2}^{2}\tanh r\sinh \lambda }{\sqrt{p_{cd}}\cosh r}%
.
\end{eqnarray*}%
Not surprisingly, the input TMSVS becomes non-Gaussian after the catalysis.
From Eq.(\ref{1-5}), I find that the LQC-TMSVS $\left\vert \psi
_{LQC}\right\rangle _{ab}$\ is actually a superposition state of $%
S_{2}\left( \lambda \right) \left\vert 0_{a},0_{b}\right\rangle $, $a^{\dag
}b^{\dag }S_{2}\left( \lambda \right) \left\vert 0_{a},0_{b}\right\rangle $,
and $a^{\dag 2}\allowbreak b^{\dag 2}S_{2}\left( \lambda \right) \left\vert
0_{a},0_{b}\right\rangle $ with a certain ratio. Note that the coefficients $%
c_{0}$, $c_{1}$, $c_{2}$,\ and $\lambda $ are all the functions of the input
squeezing parameter $r$ and the transmission coefficients $t_{1},t_{2}$ of
the BSs. Meanwhile, this state can also be looked at as a non-Gaussian state
by operating coherent superposition operator $\left( c_{0}+c_{1}a^{\dag
}b^{\dag }+c_{2}a^{\dag 2}\allowbreak b^{\dag 2}\right) $ on $S_{2}\left(
\lambda \right) \left\vert 0_{a},0_{b}\right\rangle $. So, I conclude that
local quantum-optical catalysis operation plays a role of preparing the
non-Gaussian entangled states. In the limit of $t_{1}=t_{2}=1$, $\left\vert
\psi _{LQC}\right\rangle _{ab}\rightarrow \left\vert \psi _{0}\right\rangle
_{ab}$, i.e., the output state is just the input one. While at least one of $%
t_{1},t_{2}$ is zero, leading to $\lambda =0,$ $c_{0}=0,$ $c_{1}=1,$ $%
c_{2}=0,$ so $\left\vert \psi _{LQC}\right\rangle _{ab}\rightarrow
\left\vert 1_{a},1_{b}\right\rangle $, i.e., the output state is a twin
single-photon Fock state.

By the way, I often use the catalysis parameters $T_{j}=t_{j}^{2}$ $(j=1,2)$
(i.e., the transmittance for each BS) in my following discussion and
analysis. Compared with the input TMSVS, what optimal properties will emerge
for the LQC-TMSVS? By tuning the input and catalysis parameters of the
interaction, the LQC-TMSVS may be modulated, generating a wide range of
entanglement phenomena, as I show in the next sections.

\subsection{Success probability of detection}

Normalization is important for discussing the properties of a quantum state.
The normalization factor of the LQC-TMSVS in theory is actually the
probability $p_{cd}$ of detecting successfully single photon at the modes $c$
and $d$ in experiment. The density operator of the LQC-TMSVS $\rho
_{LQC}=\left\vert \psi _{LQC}\right\rangle _{ab}\left\langle \psi
_{LQC}\right\vert $ is expressed in Appendix B. According to $\mathrm{Tr}%
\left( \rho _{LQC}\right) =1$, the success probability to get $\left\vert
\psi _{LQC}\right\rangle _{ab}$\ from my proposal is given by%
\begin{eqnarray}
p_{cd} &=&p_{0}(a_{0}+a_{1}\tanh ^{2}r+a_{2}\tanh ^{4}r  \notag \\
&&+a_{3}\tanh ^{6}r+a_{4}\tanh ^{8}r),  \label{1-6}
\end{eqnarray}%
with $p_{0}=\cosh ^{10}\lambda /\cosh ^{2}r$ and

\bigskip

\begin{eqnarray*}
a_{0} &=&t_{1}^{2}t_{2}^{2}, \\
a_{1} &=&1-4t_{1}^{2}+4t_{1}^{4}-4t_{2}^{2}+4t_{2}^{4}+16t_{1}^{2}t_{2}^{2}
\\
&&-16t_{1}^{4}t_{2}^{2}-16t_{1}^{2}t_{2}^{4}+11t_{1}^{4}t_{2}^{4}, \\
a_{2}
&=&11t_{1}^{2}t_{2}^{2}-28t_{1}^{4}t_{2}^{2}-28t_{1}^{2}t_{2}^{4}+64t_{1}^{4}t_{2}^{4}+16t_{1}^{6}t_{2}^{2}
\\
&&+16t_{1}^{2}t_{2}^{6}-28t_{1}^{4}t_{2}^{6}-28t_{1}^{6}t_{2}^{4}+11t_{1}^{6}t_{2}^{6},
\\
a_{3}
&=&11t_{1}^{4}t_{2}^{4}-16t_{1}^{6}t_{2}^{4}-16t_{1}^{4}t_{2}^{6}+4t_{1}^{8}t_{2}^{4}+4t_{1}^{4}t_{2}^{8}
\\
&&+16t_{1}^{6}t_{2}^{6}-4t_{1}^{8}t_{2}^{6}-4t_{1}^{6}t_{2}^{8}+t_{1}^{8}t_{2}^{8},
\\
a_{4} &=&t_{1}^{6}t_{2}^{6}.
\end{eqnarray*}

In Fig. 3, I plot the distribution of the success probability $p_{cd}$\ in ($%
T_{1},T_{2}$) space for $r=0.5$\ and in ($r,T$) space for the symmetric
catalysis $T_{1}=T_{2}=T$. It is found that the detection probability of
success is relatively low for the case of low transmissivity. The maximum
success probability is 1 for the limit case of $T_{1}=T_{2}=1$. While at
least one of $T_{1},T_{2}$ is zero, I find that $p_{cd}\rightarrow \left(
1-2T_{j}\right) ^{2}\tanh ^{2}r/\cosh ^{2}r$ $\left( j=1,2\right) $.

\begin{figure}[tbp]
\label{Fig3} \centering\includegraphics[width=1.0\columnwidth]{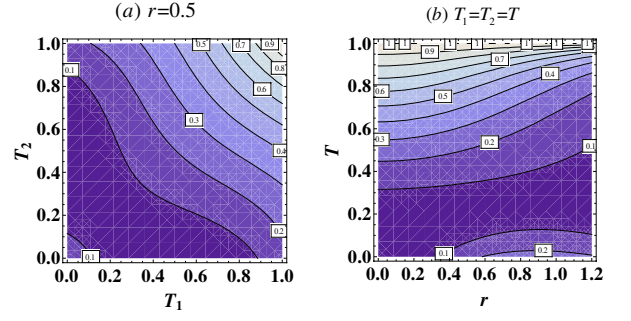}
\caption{(Color online) Success probability $p_{cd}$ of detection as a
function of the input parameter $r$ and the catalysis parameter $T_{1},T_{2}$%
. (a) In ($T_{1},T_{2}$) space for $r=0.5$; (b) in ($r,T$) space.}
\end{figure}

\section{Entanglement properties}

In contrast with the input TMSVS, can the local quantum-optical catalysis be
useful to enhance the entanglement properties? If possible, then how can I
adjust the catalysis parameters in the process of preparing the LQC-TMSVS?
In this section, I shall discuss the entanglement properties of the
LQC-TMSVS quantified by the von Neumann entropy and the EPR correlation.

\subsection{Degree of entanglement}

Entanglement for a pure state in Schmidt form, $\left\vert \psi
\right\rangle _{ab}=\sum_{n}\omega _{n}\left\vert \alpha _{n}\right\rangle
_{a}\left\vert \beta _{n}\right\rangle _{b}$ ($\omega _{n}$ real positive),
with the orthonormal states $\left\vert \alpha _{n}\right\rangle _{a}$ and $%
\left\vert \beta _{n}\right\rangle _{b}$, is quantified by the partial von
Neumann entropy of the reduced density operator,\ i.e.,%
\begin{equation}
E\left( \left\vert \psi \right\rangle _{ab}\right) =-\mathrm{Tr}\left( \rho
_{a}\log _{2}\rho _{a}\right) =-\sum_{n}\omega _{n}^{2}\log _{2}\omega
_{n}^{2},  \label{2-1}
\end{equation}%
where the local state is given by $\rho _{a}=\mathrm{Tr}_{b}\left(
\left\vert \psi \right\rangle _{ab}\left\langle \psi \right\vert \right) $ 
\cite{31}. The LQC-TMSVS $\left\vert \psi _{LQC}\right\rangle _{ab}$ written
in Schmidt form yields%
\begin{equation}
\left\vert \psi _{LQC}\right\rangle _{ab}=\sum_{n=0}^{\infty }\omega
_{n}\left\vert n_{a},n_{b}\right\rangle ,  \label{2-2}
\end{equation}%
where the Schmidt coefficient is given by%
\begin{equation}
\omega _{n}=\frac{\left( t_{1}^{2}-n+nt_{1}^{2}\right) \left(
t_{2}^{2}-n+nt_{2}^{2}\right) \left( t_{1}t_{2}\right) ^{n-1}\tanh ^{n}r}{%
\sqrt{p_{cd}}\cosh r}.  \label{2-3}
\end{equation}%
The entanglement amount of the LQC-TMSVS $E\left( \left\vert \psi
_{LQC}\right\rangle _{ab}\right) $ can be evaluated numerically by these
Schmidt coefficients, as shown in Fig. 4.

In the limit cases, when at least one of $t_{1}$ or $t_{2}$ is zero, the
output state corresponding to $\left\vert 1_{a},1_{b}\right\rangle $ is
separate, $E=0$. While $t_{1}=t_{2}=1$, leading to $\omega _{n}^{2}=\tanh
^{2n}r/\cosh ^{2}r$, the output state is just the TMSVS (the input state),
whose amount of entanglement is analytically given by \cite{32,33} 
\begin{equation}
E\left( \left\vert \psi _{0}\right\rangle _{ab}\right) =\cosh ^{2}r\log
_{2}\cosh ^{2}r-\sinh ^{2}r\log _{2}\sinh ^{2}r.  \label{2-8}
\end{equation}%
\begin{figure}[tbp]
\label{Fig4} \centering\includegraphics[width=1.0\columnwidth]{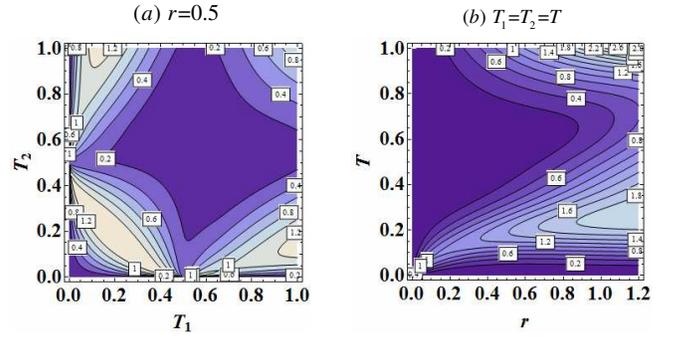}
\caption{(Color online) Neumann entropy $E$ as a function of the input
parameter $r$ and the catalysis parameter $T_{1},T_{2}$. (a) In ($T_{1},T_{2}
$) space for $r=0.5$; (b) in ($r,T$) space. }
\end{figure}

In order to understand whether the entanglement is enhanced by local
quantum-optical catalysis, I compare the von Neumann entropy values of the
LQC-TMSVSs with the TMSVSs. If $E\left( \left\vert \psi _{LQC}\right\rangle
_{ab}\right) >E\left( \left\vert \psi _{0}\right\rangle _{ab}\right) $, then
the entanglement is enhanced in principle, or else it is weakened.

There are three feasibility regions having $E\left( \left\vert \psi
_{LQC}\right\rangle _{ab}\right) >E\left( \left\vert \psi _{0}\right\rangle
_{ab}\right) $, as shown in Fig.5. One region is located in the low
transmissivities of two BSs, i.e., $T_{1},T_{2}\in (0,0.5)$, the other two
ones are located in one-small--one-large transmissivity of two BSs, i.e., $%
T_{1}\in (0,0.5)\ $but $T_{2}\in (0.5,1)$ or $T_{2}\in (0,0.5)\ $but $%
T_{1}\in (0.5,1)$. Three sections of Fig.5 with $r=0.02,0.2,0.7$ are
reshaped in Fig.6. With increasing the input parameter $r$, the enhancement
region decreases and disappears at threshold $r=0.785$, as shown in Fig.6. 
\begin{figure}[tbp]
\label{Fig5} \centering\includegraphics[width=0.9\columnwidth]{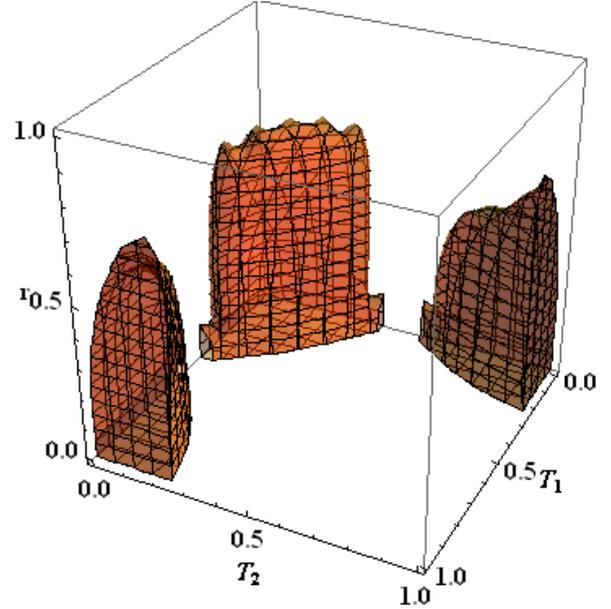}
\caption{(Color online) Three dimensional plot of the feasibility region for 
$E\left( \left\vert \protect\psi _{LQC}\right\rangle _{ab}\right) >E\left(
\left\vert \protect\psi _{0}\right\rangle _{ab}\right) $ in ($r,T_{1},T_{2}$%
) space.}
\end{figure}
\begin{figure}[tbp]
\label{Fig6} \centering\includegraphics[width=1.0\columnwidth]{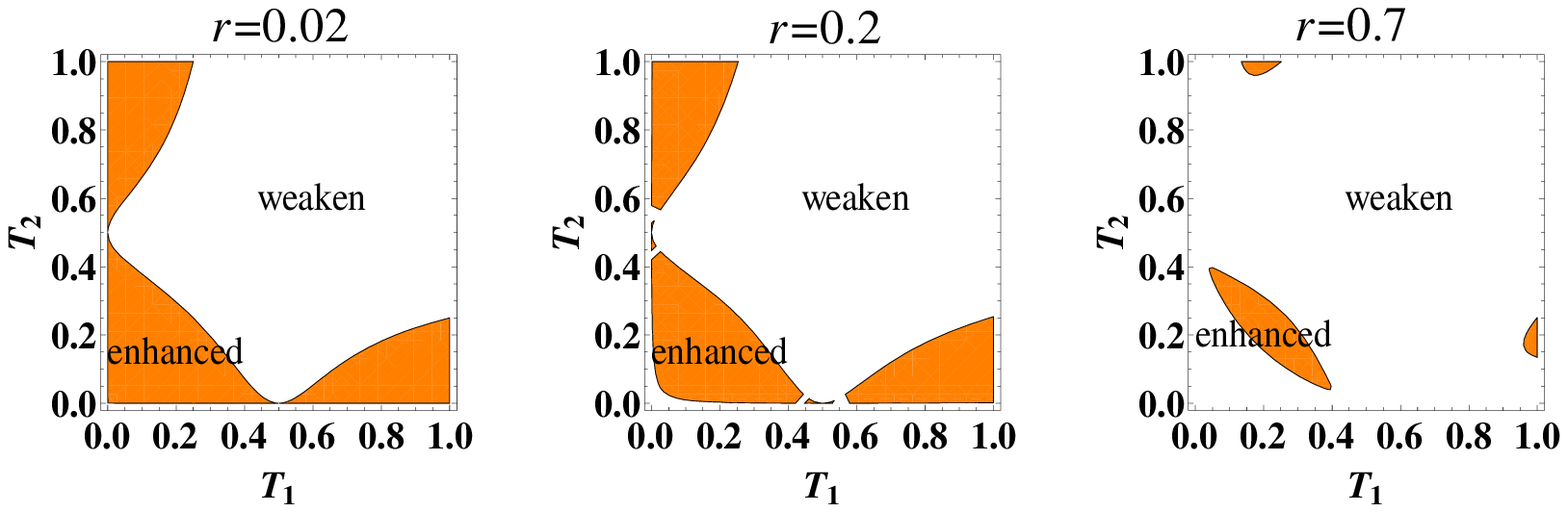}
\caption{(Color online) Plot of the feasibility region for $E\left(
\left\vert \protect\psi _{LQC}\right\rangle _{ab}\right) >E\left( \left\vert 
\protect\psi _{0}\right\rangle _{ab}\right) $ in ($T_{1},T_{2}$) space with
different $r=0.02,0.2,0.7$, also three sections of Fig. 5. If $r$ is lager
than a threshold value $0.785$, the enhancement is impossible.}
\end{figure}

Next, I discuss the symmetric catalysis case, i.e., assuming $T=T_{1}=T_{2}$%
. The feasibility region for enhancing the entanglement is depicted in the ($%
r,T$) plain space in Fig.7. The enhancement happens in small-squeezing ($%
0<r<0.785$) and low transmissivity ($0<T<0.25$) regimes. In Figs. 8. (a) I
plot the von Neumann entropy $E\left( \left\vert \psi _{LQC}\right\rangle
_{ab}\right) $ as a function of the input squeezing parameter $r$ for
different $T=0.1,0.3$, compared with $T=1$ (corresponding to the input
TMSVS). With reference to the curve of the TMSVS, one sees that the
enhancement is possible for $T=0.1$ but not for $T=0.3$. Compared with the
corresponding TMSVSs (the red dashed line), I plot the von Neumann entropy $%
E\left( \left\vert \psi _{LQC}\right\rangle _{ab}\right) $ as a function of
the catalysis parameter $T$ for different input parameter $r$ $=0.2,0.785,0.9
$ in Figs. 8. (b)-(d). For instance, when $r$ $=0.2$, the enhancement of
entanglement will happen in a certain catalysis range (about $\left(
0.03,0.23\right) $) [see Fig.8 (b)]. But, above the threshold value $r$ $%
=0.785$, the enhancement is impossible, as shown in Fig.8 (d) for $r$ $=0.9$%
. 
\begin{figure}[tbp]
\label{Fig7} \centering\includegraphics[width=0.9\columnwidth]{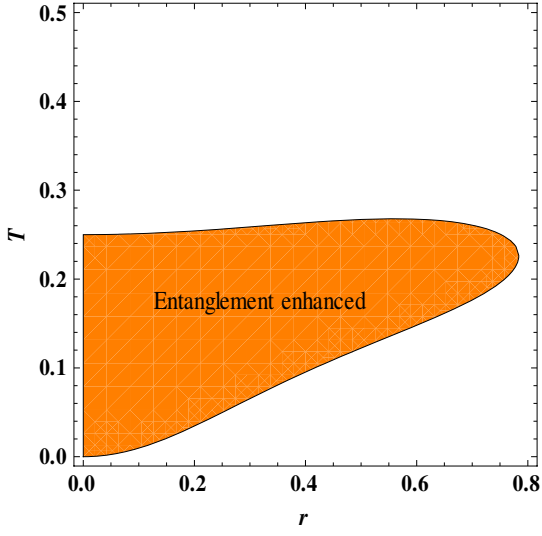}
\caption{(Color online) Plot of the feasibility region for enhancing
entanglement, that is $E\left( \left\vert \protect\psi _{LQC}\right\rangle
_{ab}\right) >E\left( \left\vert \protect\psi _{0}\right\rangle _{ab}\right) 
$ in ($r,T$) space. }
\end{figure}
\begin{figure}[tbp]
\label{Fig8} \centering\includegraphics[width=1.0\columnwidth]{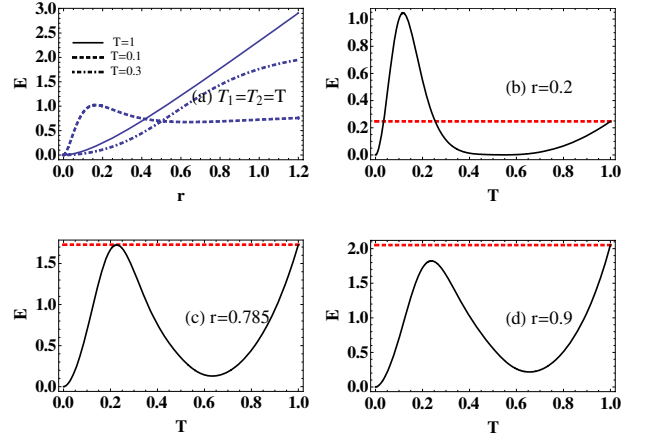}
\caption{(Color online) Neumann entropy $E\left( \left\vert \protect\psi %
_{LQC}\right\rangle _{ab}\right) $ are depicted in (a) as a function of the
input squeezing parameter $r$ for different $T=0.1,0.3$, compared with $T=1$
(corresponding to the input TMSVS); in (b)-(d) as a function of the
catalysis parameter $T$ (black solid line) for (b) $r$ $=0.2$, (c) $r$ $%
=0.785$, (d) $r$ $=0.9$, respectively, compared with their TMSVSs (red
dashed line). }
\end{figure}

From the above discussion, I conclude that the degree of entanglement
measured by the von Neumann entropy turns out to be enhanced only in the
small-squeezing and low-transmissivity parameter spaces.

\subsection{Second-order Einstein-Podolsky-Rosen correlation}

For two-mode Gaussian entangled states, the entanglement can be fully
described by the second-order Einstein-Podolsky-Rosen (EPR) correlation,
which is characterized up to second-order moments of the state \cite%
{34,35,36}. For two-mode non-Gaussian entangled states, however, the
entanglement is fully described with all orders of moments \cite{37,38}. It
is known to all that a TMSVS (Gaussian) is the correlated state of two field
modes $a$ and $b$ (signal and idle) that can be generated by a nonlinear
medium \cite{39}. But, after operating local quantum-optical catalysis on
the TMSVS, how does the EPR correlation change? Here, I further investigate
the EPR correlation, another entanglement property for the LQC-TMSVS.

The EPR correlation of two-mode states is the total variance of a pair of
EPR-type operators,%
\begin{eqnarray}
\mathrm{EPR} &=&\Delta ^{2}\left( x_{a}-x_{b}\right) +\Delta ^{2}\left(
p_{a}+p_{b}\right)   \notag \\
&=&2\left( 1+\left\langle a^{\dag }\allowbreak a\right\rangle +\left\langle
b^{\dag }\allowbreak b\right\rangle -\left\langle a^{\dag }\allowbreak
b^{\dag }\right\rangle -\left\langle a\allowbreak b\right\rangle \right)  
\notag \\
&&-2\left( \left\langle a\right\rangle -\left\langle b^{\dag }\allowbreak
\right\rangle \right) \left( \left\langle a^{\dag }\right\rangle
-\left\langle b\allowbreak \right\rangle \right) ,  \label{2-4}
\end{eqnarray}%
where $x_{j}=\frac{1}{\sqrt{2}}\left( j+j^{\dag }\right) $ and $p_{j}=\frac{%
-i}{\sqrt{2}}\left( j-j^{\dag }\right) $ ($j=a,b$). For separable two-mode
states or any classical two-mode states, the total variance is larger than
or equal to 2 \cite{34}. The condition $\mathrm{EPR}<2$, indicating quantum
entanglement, can be an important resource in continuous variable quantum
information processing protocols.

Given a LQC-TMSVS, one can evaluate the EPR\ correlation with the
expectation values in Eq. (\ref{2-4}). Using the general expression of $%
\left\langle a^{\dag k_{1}}\allowbreak b^{\dag k_{2}}a^{l_{1}}\allowbreak
b^{l_{2}}\right\rangle $ in Appendix D, I prove that $\left\langle a^{\dag
}\allowbreak \right\rangle =\left\langle \allowbreak b^{\dag }\right\rangle
=\left\langle a\right\rangle =\left\langle \allowbreak b\right\rangle =0$ and%
\begin{eqnarray}
\left\langle a^{\dag }\allowbreak a\right\rangle  &=&M(x_{0}+x_{1}\tanh
r+x_{2}\tanh ^{2}r+x_{3}\tanh ^{3}r  \notag \\
&&+x_{4}\tanh ^{4}r+x_{5}\tanh ^{5}r+x_{6}\tanh ^{6}r  \notag \\
&&+x_{7}\tanh ^{7}r+x_{8}\tanh ^{8}r+x_{9}\tanh ^{9}r),  \label{2-5}
\end{eqnarray}%
\begin{eqnarray}
\left\langle b^{\dag }\allowbreak b\right\rangle  &=&M(y_{0}+y_{1}\tanh
r+y_{2}\tanh ^{2}r+y_{3}\tanh ^{3}r  \notag \\
&&+y_{4}\tanh ^{4}r+y_{5}\tanh ^{5}r+y_{6}\tanh ^{6}r  \notag \\
&&+y_{7}\tanh ^{7}r+y_{8}\tanh ^{8}r+y_{9}\tanh ^{9}r),  \label{2-6}
\end{eqnarray}%
as well as%
\begin{eqnarray}
\left\langle a^{\dag }\allowbreak b^{\dag }\right\rangle  &=&\left\langle
a\allowbreak b\right\rangle =N(z_{0}+z_{1}\tanh r+z_{2}\tanh ^{2}r  \notag \\
&&+z_{3}\tanh ^{3}r+z_{4}\tanh ^{4}r+z_{5}\tanh ^{5}r  \notag \\
&&+z_{6}\tanh ^{6}r+z_{7}\tanh ^{7}r+z_{8}\tanh ^{8}r),  \label{2-7}
\end{eqnarray}%
where I have set $x_{i}$, $y_{i}$, $z_{i}$ in Appendix E and 
\begin{eqnarray*}
M &=&(\cosh ^{12}\lambda \tanh r)/(p_{cd}\cosh ^{2}r), \\
N &=&(\cosh ^{12}\lambda \tanh \lambda )/(p_{cd}\cosh ^{2}r).
\end{eqnarray*}%
Upon substituting the above equations into Eq. (\ref{2-4}), the EPR
correlation of the LQC-TMSVS $\mathrm{EPR}\left( \left\vert \psi
_{LQC}\right\rangle _{ab}\right) $ can be calculated explicitly, which
depends on the input squeezing degree $r$ and the catalysis parameters $%
T_{1},T_{2}$. In the limit of $T_{1}=T_{2}=1$, $\mathrm{EPR}\left(
\left\vert \psi _{LQC}\right\rangle _{ab}\right) $ reduces to $\mathrm{EPR}%
\left( \left\vert \psi _{0}\right\rangle _{ab}\right) =2e^{-2r}$, which
tends to zero asymptotically for $r\rightarrow \infty $. In Fig.9, I plot
the EPR correlation of the LQC-TMSVS in ($T_{1},T_{2}$) space for $r=0.5$
and in ($r,T$) space under the condition $\mathrm{EPR}<2$. One can see that
there exists a threshold curve (boundary of $\mathrm{EPR}=2$) as a function
of $T_{1}$ and $T_{2}$ for $r=0.5$ in Fig. 9(a) and as a function of $r$ and 
$T$ in Fig. 9(b), respectively. 
\begin{figure}[tbp]
\label{Fig9} \centering\includegraphics[width=1.0\columnwidth]{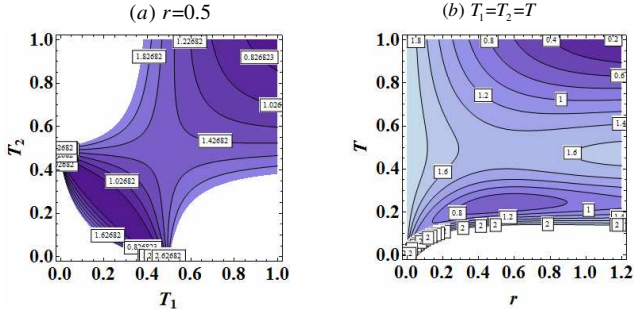}
\caption{(Color online) EPR correlation as a function of the input squeezing
parameter $r$ and the catalysis parameter $T_{1},T_{2}$. (a) In ($T_{1},T_{2}
$) space for $r=0.5$; (b) in ($r,T$) space. The Colored region represents
the condition $\mathrm{EPR}<2$.}
\end{figure}

To exhibit whether the EPR correlation is enhanced, the fact that $\mathrm{%
EPR}\left( \left\vert \psi _{LQC}\right\rangle _{ab}\right) $ must be
smaller than $\mathrm{EPR}\left( \left\vert \psi _{0}\right\rangle
_{ab}\right) $ must hold. The feasibility enhancement region of the EPR
correlation is shown in Fig. 10. Three sections of Fig. 10 are shown in Fig.
11. Obviously, the enhancement happens only in one region with
small-squeezing and low-transmissivity, unlike that of the degree of
entanglement in Figs. 5 and 6. Moreover, with increasing the input parameter 
$r$, the enhancement region decreases and disappears at threshold $r=0.585$. 
\begin{figure}[tbp]
\label{Fig10} \centering\includegraphics[width=0.9\columnwidth]{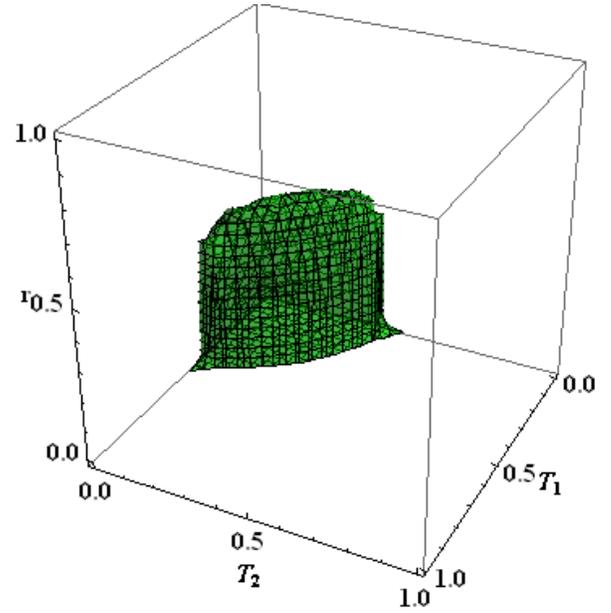}
\caption{(Color online) Three-dimensional plot of the feasibility region for
enhancing EPR correlation, that is, $\mathrm{EPR}\left( \left\vert \protect%
\psi _{LQC}\right\rangle _{ab}\right) <\mathrm{EPR}\left( \left\vert \protect%
\psi _{0}\right\rangle _{ab}\right) $, in ($r,T_{1},T_{2}$) space. }
\end{figure}
\begin{figure}[tbp]
\label{Fig11} \centering\includegraphics[width=1.0\columnwidth]{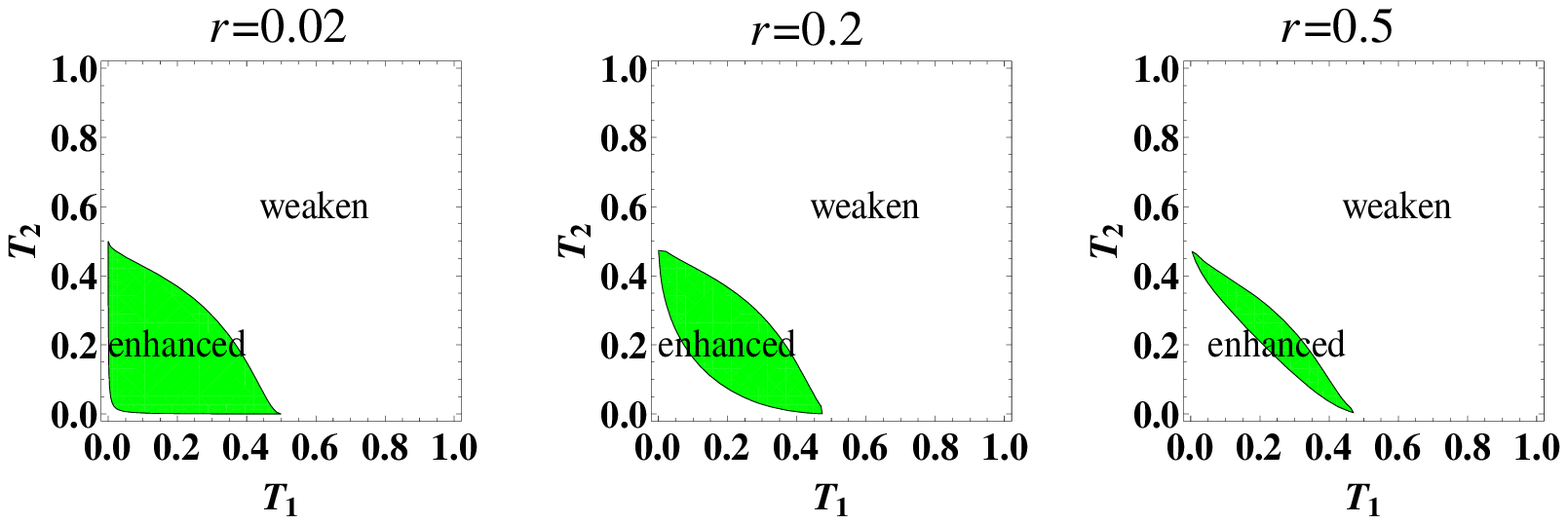}
\caption{(Color online) Plot of the feasibility region for enhancing EPR
correlation, that is $\mathrm{EPR}\left( \left\vert \protect\psi %
_{LQC}\right\rangle _{ab}\right) <\mathrm{EPR}\left( \left\vert \protect\psi %
_{0}\right\rangle _{ab}\right) $, in ($T_{1},T_{2}$) space with different $%
r=0.02,0.2,0.5$, also three sections of Fig.10. If $r$ is larger than a
threshold value $0.585$, the enhancement is impossible.}
\end{figure}

The feasibility region for enhancing the EPR correlation is depicted in the (%
$r,T$) plain space in Fig. 12. For a small-squeezing ($0<r<0.585$) and
low-transmissivity ($0<T<0.3$), the quantum-optical catalysis enhance the
EPR correlation of the TMSVS (see Fig. 2). In Fig. 13, I plot the EPR
correlation of the LQC-TMSVS as a function of $r$ or $T$. In general, the
EPR correlation of the TMSVS\ is enhanced with the squeezing parameter $r$,
but it may not be always true\ for the case of $T=0.1$, as shown in Fig.13
(a). I particulary compare the EPR correlation of the LQC-TMSVS with that of
the TMSVS for the cases $r=0.2,0.585,0.7$ in Fig.13 (b)-(d). For a moderate
catalysis parameter $0.12<T<0.3$, the catalysis operation gives the better
EPR correlation for $r=0.2$. For a large squeezing ($r>0.585$), the
quantum-optical catalysis becomes the worse\ operation. 
\begin{figure}[tbp]
\label{Fig12} \centering\includegraphics[width=0.9\columnwidth]{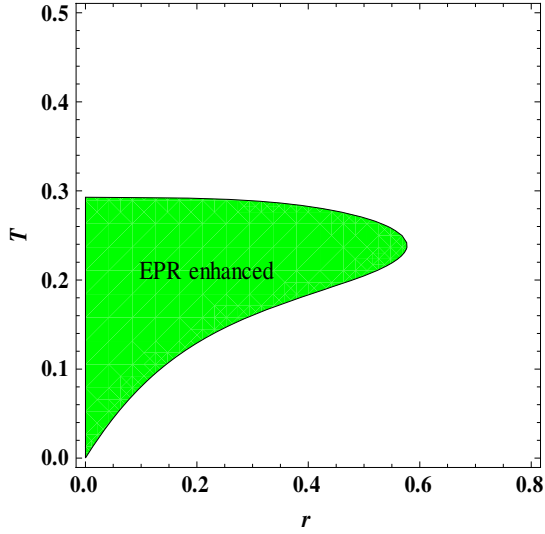}
\caption{(Color online) Plot of the feasibility region for enhancing EPR
correlation, that is $\mathrm{EPR}\left( \left\vert \protect\psi %
_{LQC}\right\rangle _{ab}\right) <\mathrm{EPR}\left( \left\vert \protect\psi %
_{0}\right\rangle _{ab}\right) $ in ($r,T$) space. }
\end{figure}
\begin{figure}[tbp]
\label{Fig13} \centering\includegraphics[width=1.0\columnwidth]{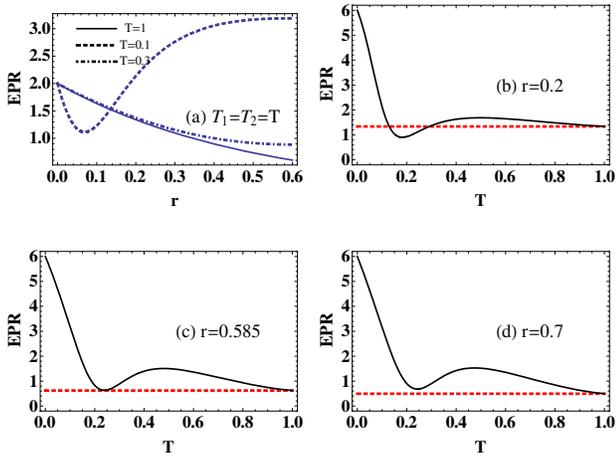}
\caption{(Color online) (a) EPR correlation as a function of the input
parameter $r$ for different $T=0.1,0.3$, compared with $T=1$ (corresponding
to the input TMSVS); in (b)-(d) as a function of $T$ for different input
parameter $r$ $=0.2,0.585,0.7$, compared with their TMSVSs (the red dashed
line). }
\end{figure}

\section{Quantum teleportation using non-Gaussian entangled state}

After employing local quantum-optical catalysis on the TMSVS, I can see that
the degree of entanglement and the EPR correlation can be enhanced in
small-squeezing and low-transmissivity parameter regime. Now I consider the
LQC-TMSVS as entangled resources in the Braunstein and Kimble (BK) protocol 
\cite{1} to teleport a coherent state $\left\vert \gamma \right\rangle $ in
CV teleportation. The fidelity between an input state and the output state
is usually used as a measure to describe the quality of the quantum
teleportation (QT).

For a CV system, a teleportation scheme has been proposed according to the
characteristic functions (CFs) of the quantum states concluding input,
source and teleported states \cite{40}. For the input coherent state, it is
sufficient to calculate the teleportation fidelity for a particular input
coherent state since there is no difference between the amplitudes of the
input and output coherent states in the BK protocol. For brevity I take $%
\gamma =0$, and then I only calculate the fidelity of teleporting the input
vacuum state with the CF $\chi _{in}(z)=\exp [-|z|^{2}/2]$. The CF of the
LQC-TMSVS (entangled resource or channel) $\left\vert \psi
_{LQC}\right\rangle _{ab}$\ is given by%
\begin{equation}
\chi _{E}\left( \alpha ,\beta \right) =\mathrm{Tr}\left( D_{a}\left( \alpha
\right) D_{b}\left( \beta \right) \rho _{LQC}\right) ,  \label{3-1}
\end{equation}%
where $D_{a}\left( \alpha \right) =e^{\alpha a^{\dag }-\alpha ^{\ast
}a},D_{b}\left( \beta \right) =$\ $e^{\beta b^{\dag }-\beta ^{\ast }b}$ are
the displacement operators. The detailed calculation procedure and result of 
$\chi _{E}\left( \alpha ,\beta \right) $ are shown in Appendix F. The CF $%
\chi _{out}(z)$ of the output state can be related to the CFs of the input
state and the entangled source by formula $\chi _{out}(z)=\chi _{in}(z)\chi
_{E}\left( z^{\ast },z\right) $. Hence the fidelity of QT of CVs can be
obtained as \cite{41}%
\begin{equation}
F=\int \frac{d^{2}z}{\pi }\chi _{in}(-z)\chi _{out}(z).  \label{3-2}
\end{equation}%
Thus $F$ yields%
\begin{eqnarray}
F &=&\frac{p_{0}}{4p_{cd}}(m_{0}+m_{1}\tanh r+m_{2}\tanh ^{2}r  \notag \\
&&+m_{3}\tanh ^{3}r+m_{4}\tanh ^{4}r),  \label{3-3}
\end{eqnarray}%
where%
\begin{eqnarray*}
m_{0} &=&2t_{1}^{2}t_{2}^{2}, \\
m_{1} &=&2t_{1}t_{2}-4t_{1}^{3}t_{2}-4t_{1}t_{2}^{3}-2t_{1}^{3}t_{2}^{3}, \\
m_{2} &=&1-4t_{1}^{2}+4t_{1}^{4}-4t_{2}^{2}+4t_{2}^{4}+10t_{1}^{2}t_{2}^{2}
\\
&&-2t_{1}^{4}t_{2}^{2}-2t_{1}^{2}t_{2}^{4}+5t_{1}^{4}t_{2}^{4}, \\
m_{3}
&=&t_{1}t_{2}-t_{1}^{3}t_{2}-2t_{1}^{5}t_{2}-t_{1}t_{2}^{3}-2t_{1}t_{2}^{5}
\\
&&-2t_{1}^{3}t_{2}^{3}+t_{1}^{5}t_{2}^{3}+t_{1}^{3}t_{2}^{5}-3t_{1}^{5}t_{2}^{5},
\\
m_{4}
&=&t_{1}^{2}t_{2}^{2}-t_{1}^{4}t_{2}^{2}+t_{1}^{6}t_{2}^{2}-t_{1}^{2}t_{2}^{4}+t_{1}^{2}t_{2}^{6}
\\
&&+2t_{1}^{4}t_{2}^{4}-t_{1}^{6}t_{2}^{4}-t_{1}^{4}t_{2}^{6}+t_{1}^{6}t_{2}^{6}.
\end{eqnarray*}%
In the limit case of $t_{1}^{2}=t_{2}^{2}=1$, the fidelity of LQC-TMSVS $%
F\left( \left\vert \psi _{LQC}\right\rangle _{ab}\right) $ reduces to that
of the TMSVS $F\left( \left\vert \psi _{0}\right\rangle _{ab}\right)
=(1+\tanh r)/2$, which is $0.5$ for $r=0$ and tends to $1$ asymptotically
for $r\rightarrow \infty $. In Fig.14, I show the fidelity of teleporting a
coherent state using the resource (LQC-TMSVS) in ($T_{1},T_{2}$) space for $%
r=0.5$ and in ($r,T$) space. The red line denotes the boundary with $F=0.5$.
The fidelity over the classical limit 0.5 may be considered as a successful
quantum protocol \cite{42}. 
\begin{figure}[tbp]
\label{Fig14} \centering\includegraphics[width=1.0\columnwidth]{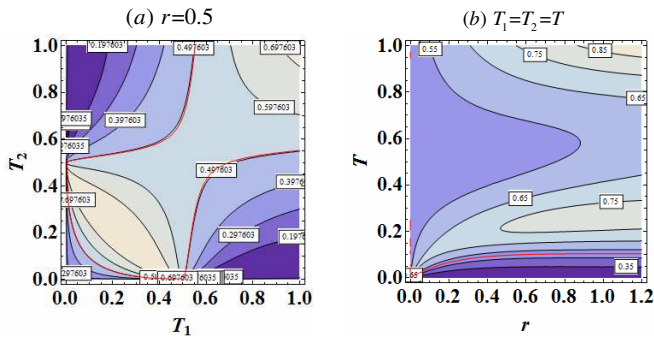}
\caption{(Color online) Teleportation fidelity of a coherent state with the
LQC-TMSVS as a function of the input squeezing parameter $r$ and the
catalysis parameter $T_{1},T_{2}$. (a) In ($T_{1},T_{2}$) space for $r=0.5$;
(b) in ($r,T$) space. The red line is the boundary with $F=0.5$.}
\end{figure}

Similar analysis of the teleportation fidelity is performed like that of the
degree of entanglement and the EPR correlation in Sec. III. In Fig. 15, I
plot the feasibility region for enhancing teleportation fidelity of a
coherent state with the LQC-TMSVS, i.e., $F\left( \left\vert \psi
_{LQC}\right\rangle _{ab}\right) >F\left( \left\vert \psi _{0}\right\rangle
_{ab}\right) $, in ($r,T_{1},T_{2}$) space. Figures in Fig. 16 are three
sections of Fig. 15 with $r=002,0.2,0.5$. In Fig. 17, I display the
feasibility region in ($r,T$) space for enhancing teleportation fidelity of
a coherent state using the LQC-TMSVS. The teleportation fidelity as a
function of $r$ or $T$ is plotted in Fig. 18. Compared with the TMSVS as the
entangled resource, the enhancement of the teleportation fidelity is found
in the range of $0<r<0.6$ and $0<T<0.27$. All these figures indicate that
local quantum-optical catalysis can enhance the teleportation fidelity at
the small-squeezing and low-transmissivity parameter regime. 
\begin{figure}[tbp]
\label{Fig15} \centering\includegraphics[width=0.9\columnwidth]{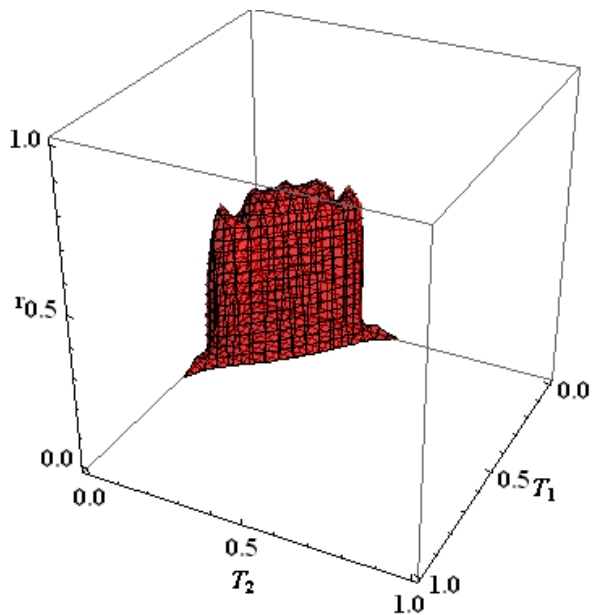}
\caption{(Color online) Three-dimensional plot of the feasibility region for
enhancing teleportation fidelity of a coherent state with the LQC-TMSVS,
that is, $F\left( \left\vert \protect\psi _{LQC}\right\rangle _{ab}\right)
>F\left( \left\vert \protect\psi _{0}\right\rangle _{ab}\right) $, in ($%
r,T_{1},T_{2}$) space. }
\end{figure}
\begin{figure}[tbp]
\label{Fig16} \centering\includegraphics[width=1.0\columnwidth]{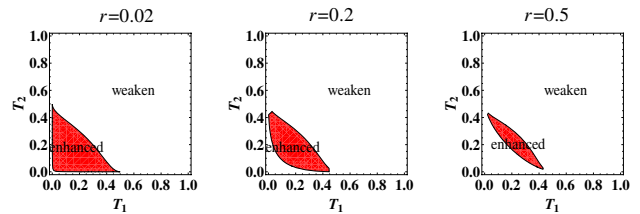}
\caption{(Color online) Plot of the feasibility region for enhancing
teleportation fidelity of a coherent state with the LQC-TMSVS, that is $%
F\left( \left\vert \protect\psi _{LQC}\right\rangle _{ab}\right) >F\left(
\left\vert \protect\psi _{0}\right\rangle _{ab}\right) $ in ($T_{1},T_{2}$)
space with different $r=0.02,0.2,0.5$, also three sections of Fig. 15. If $r$
is larger than a threshold value $0.6$, the enhancement is impossible.}
\end{figure}
\begin{figure}[tbp]
\label{Fig17} \centering\includegraphics[width=0.9\columnwidth]{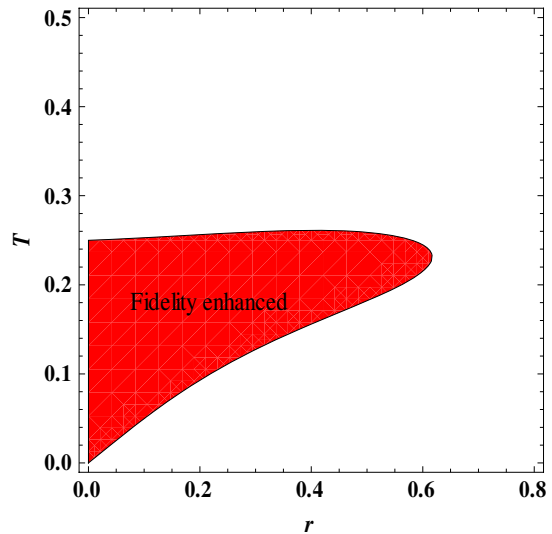}
\caption{(Color online) Plot of the feasibility region for enhancing
teleportation fidelity of a coherent state with the LQC-TMSVS, that is, $%
F\left( \left\vert \protect\psi _{LQC}\right\rangle _{ab}\right) >F\left(
\left\vert \protect\psi _{0}\right\rangle _{ab}\right) $ in ($r,T$) space. }
\end{figure}
\begin{figure}[tbp]
\label{Fig18} \centering\includegraphics[width=1.0\columnwidth]{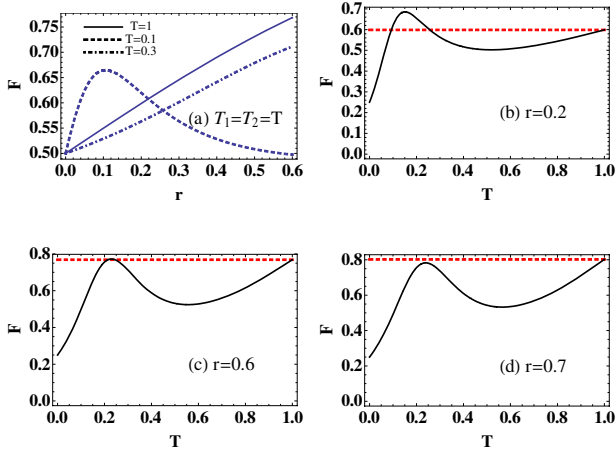}
\caption{(Colour online) (a) Teleportation fidelity of a coherent state with
the LQC-TMSVS as a function of the input parameter $r$ for different $%
T=0.1,0.3$, compared with $T=1$ (corresponding to the input TMSVS); in
(b)-(d) as a function of $T$ for different input parameter $r$ $=0.2,0.6,0.7$%
, compared with their TMSVSs (the red dashed line). }
\end{figure}

\section{Discussion and Conclusion}

Interestingly, when comparing the different enhancement feasibility regions
of the quantities [degree of entanglement (orange), the EPR correlation
(green), and the teleportation fidelity (red)] of the LQC-TMSVS in Figs. 5,
10, and 15 and 7, 12, and 17, I find that these enhancement regions do not
overlap completely and locate in different input and catalysis parameter
intervals. Taking the symmetric catalysis as example, the enhancement
regions are different as (i) $0<r<0.785$ and $0<T<0.25$\ for the degree of
entanglement (ii) $0<r<0.585$ and $0<T<0.3$\ for the EPR correlation; (iii) $%
0<r<0.6$ and $0<T<0.27$\ for the teleportation fidelity. I further reshape
each two of the three plots (Figs. 7, 12, and 17) in the same graph, as
shown in Fig.19. The conclusions are concluded by answering the following
question: If A is enhanced, then must B be enhanced?, as demonstrated in
Table I. For instance, there exists a parameter region where there is no EPR
correlation enhancement, nevertheless, the fidelity enhancement is achieved
[see the red area in Fig.19 (f)], so my answer is \textquotedblleft
no.\textquotedblright\ For all these three quantities, there are common
enhancement region as shown in Fig.20. This region locates in the regime of
the relatively low beam-splitter transmissivities $T_{1}$\ and $T_{2}$\
(from 0 to around 0.25) and the small squeezing parameters (from 0 to around
0.6), which are the most experimentally accessible.

As the quantum-optical catalysis is an operation based on post selection,
the probability of success is naturally an issue. However, it is
disadvantageous to see that the success probabilities in the most desirable
parameter ranges (i.e., the enhancement regions) are relatively low in my
protocol, as showed in Fig.3. There is a fundamental trade-off between
success probability of the operation and the resultant enhancement in
entanglement. In addition, the catalysis operation maximizes entanglement at
low but nonzero probability. Thus, the success of detecting the single
photon, also the key of the quantum-optical catalysis, is determined by the
perfection of the detectors. As long as the detector is enough perfect, the
single photon can be detected successfully. Using the current detection
technology, the problem of low detection probability is possible to solve.
This is good news. For example, the single photon can be counted by using
superconducting single-photon detector with high efficiency ($>90\%$),
ultralow noise ($<1Hz$), and low timing jitter ($<100ps$) \cite{43}. In
experiment, it is possible to count near-infrared single photon with $95\%$\
efficiency. The measured $95\%$\ system detection efficiency is consistent
with measurements and simulations of the optical elements \cite{44}. On the
other hand, the probability of success in experiment is actually the
normalization factor for a prepared quantum state in theory. From the point
of view of quantum mechanics, once the detection is succeeding, the quantum
state can be generated. 
\begin{figure}[tbp]
\label{Fig19} \centering\includegraphics[width=1.0\columnwidth]{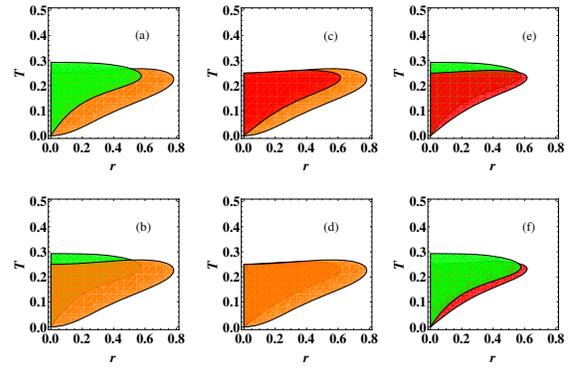}
\caption{(Color online) Comparison of the enhancing feasibility region for
each two of the three properties, i.e., the degree of entanglement (orange,
E), EPR correlation (green, EPR), and teleportation fidelity (red, F) in ($%
r,T$) space for symmetric catalysis. (a) E under EPR; (b) EPR under E; (c) E
under F; (d) F under E; (e) EPR under F; (f) F under EPR. The stack-ups
indicate the enhancement difference of these three properties. The
illustration is explained in Table I.}
\end{figure}
\begin{table}[tbp]
\caption{If A is enhanced, then must B be enhanced? }\centering%
\begin{tabular}{c||c|c|c}
\hline\hline
case & $A$ & $B$ & answer \\ \hline\hline
Fig.19(a) & $E$ & $EPR$ & no \\ \hline
Fig.19(b) & $EPR$ & $E$ & no \\ \hline
Fig.19(c) & $E$ & $F$ & no \\ \hline
Fig.19(d) & $F$ & $E$ & yes \\ \hline
Fig.19(e) & $EPR$ & $F$ & no \\ \hline
Fig.19(f) & $F$ & $EPR$ & no \\ \hline\hline
\end{tabular}%
\end{table}
\begin{figure}[tbp]
\label{Fig20} \centering\includegraphics[width=1.0\columnwidth]{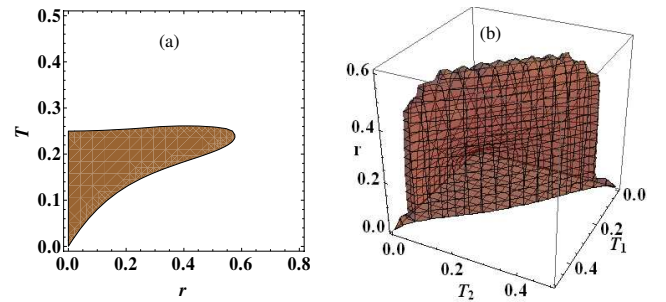}
\caption{(Colour online) The common feasibility region for enhancing
entanglement, EPR correlation and teleportation fidelity in ($r,T$) space
(a) and in ($r,T_{1},T_{2}$) space (b). The brown region are located at
small-squeezing and low transmissivities regime. }
\end{figure}

In summary, this paper presents the effects of quantum optical catalysis on
the two-mode squeezed vacuum in terms of various entanglement measures,
namely, entanglement entropy, second-order EPR correlation, and the fidelity
of quantum teleportation. The operation of the quantum-optical catalysis is
a powerful tool which can be used to increase entanglement under certain
conditions.

\begin{acknowledgments}
The author would like to thank L.-y. Hu, Z.-s. Wang, Z.-l. Duan, J.-h.
Huang, H.-c. Yuan, Q. Guo, and Y.-h. Li for helpful discussions. This work
was supported by the National Nature Science Foundation of China (Grants No.
11264018 and No. 11447002) and the Natural Science Foundation of Jiangxi
Province of China (Grants No. 20142BAB202001 and No. 20151BAB202013).
\end{acknowledgments}

\bigskip

\textbf{Appendix A: the explicit form of }$\left\vert \psi
_{LQC}\right\rangle _{ab}$

In this appendix, I derive the the explicit form of $\left\vert \psi
_{LQC}\right\rangle _{ab}$. Noting the integral form of $\left\vert \psi
_{0}\right\rangle _{ab}$,%
\begin{equation*}
\left\vert \Psi _{0}\right\rangle _{ab}=\frac{1}{\sinh r}\int \frac{%
d^{2}\alpha }{\pi }e^{-\left\vert \alpha \right\vert ^{2}\tanh ^{-1}r+\alpha
a^{\dag }+\alpha ^{\ast }b^{\dag }}\left\vert 0,0\right\rangle 
\end{equation*}%
and the differential form of Fock state $\left\vert 1\right\rangle $, such
as $\left\vert 1_{c}\right\rangle =\frac{d}{ds_{1}}e^{s_{1}c^{\dag
}}\left\vert 0_{c}\right\rangle |_{s_{1}=0}$ and $\left\vert
1_{d}\right\rangle =\frac{d}{ds_{2}}e^{s_{2}d^{\dag }}\left\vert
0_{d}\right\rangle |_{s_{2}=0}$, I rewrite $\left\vert \psi
_{LQC}\right\rangle _{ab}$\ as

\begin{eqnarray*}
&&\left\vert \psi _{LQC}\right\rangle _{ab} \\
&=&\frac{1}{\sqrt{p_{cd}}\sinh r}\frac{d^{4}}{ds_{1}ds_{2}ds_{3}ds_{4}}\int 
\frac{d^{2}\alpha }{\pi }e^{-\left\vert \alpha \right\vert ^{2}\tanh ^{-1}r}
\\
&&\left\langle 0_{c}\right\vert e^{s_{3}c}B_{1}e^{\alpha a^{\dag
}}e^{s_{1}c^{\dag }}B_{1}^{\dag }\left\vert 0_{a}\right\rangle \left\vert
0_{c}\right\rangle \text{ } \\
&&\left\langle 0_{d}\right\vert e^{s_{4}d}B_{2}e^{\alpha ^{\ast }b^{\dag
}}e^{s_{2}d^{\dag }}B_{2}^{\dag }\left\vert 0_{b}\right\rangle \left\vert
0_{d}\right\rangle |_{(s_{1},s_{2},s_{3},s_{4})=0}.
\end{eqnarray*}%
where $(s_{1},s_{2},s_{3},s_{4})=0$\ denotes $s_{1}=s_{2}=s_{3}=s_{4}=0$.
Further using the transformation in Eq.(\ref{1-3}), $B_{1}\left\vert
0_{a}\right\rangle \left\vert 0_{c}\right\rangle =\left\vert
0_{a}\right\rangle \left\vert 0_{c}\right\rangle $ and $B_{2}\left\vert
0_{b}\right\rangle \left\vert 0_{d}\right\rangle =\left\vert
0_{b}\right\rangle \left\vert 0_{d}\right\rangle $, I have%
\begin{eqnarray*}
&&\left\vert \psi _{LQC}\right\rangle _{ab} \\
&=&\frac{1}{\sqrt{p_{cd}}\sinh r}\frac{d^{4}}{ds_{1}ds_{2}ds_{3}ds_{4}}\int 
\frac{d^{2}\alpha }{\pi }e^{-\left\vert \alpha \right\vert ^{2}\tanh ^{-1}r}
\\
&&\left\langle 0_{c}\right\vert e^{s_{3}c}e^{c^{\dag }\left(
s_{1}t_{1}-\alpha r_{1}\right) }\left\vert 0_{c}\right\rangle \left\langle
0_{d}\right\vert e^{s_{4}d}e^{d^{\dag }\left( s_{2}t_{2}-\alpha ^{\ast
}r_{2}\right) }\left\vert 0_{d}\right\rangle \\
&&e^{a^{\dag }\left( \alpha t_{1}+\allowbreak s_{1}r_{1}\right) }\left\vert
0_{a}\right\rangle \otimes e^{b^{\dag }\left( s_{2}r_{2}+\alpha ^{\ast
}\allowbreak t_{2}\right) }\left\vert 0_{b}\right\rangle
|_{(s_{1},s_{2},s_{3},s_{4})=0}.
\end{eqnarray*}%
Inserting the completeness relation of coherent state $\int \frac{d^{2}z_{j}%
}{\pi }\left\vert z_{j}\right\rangle \left\langle z_{j}\right\vert =1$ $%
(j=1,2)$ in appropriate place, I have%
\begin{eqnarray*}
&&\left\vert \psi _{LQC}\right\rangle _{ab} \\
&=&\frac{1}{\sqrt{p_{cd}}\sinh r}\frac{d^{4}}{ds_{1}ds_{2}ds_{3}ds_{4}}\int 
\frac{d^{2}\alpha }{\pi }e^{-\left\vert \alpha \right\vert ^{2}\tanh ^{-1}r}
\\
&&\left\langle 0_{c}\right\vert e^{s_{3}c}\int \frac{d^{2}z_{1}}{\pi }%
\left\vert z_{1}\right\rangle \left\langle z_{1}\right\vert e^{c^{\dag
}\left( s_{1}t_{1}-\alpha r_{1}\right) }\left\vert 0_{c}\right\rangle \\
&&\left\langle 0_{d}\right\vert e^{s_{4}d}\int \frac{d^{2}z_{2}}{\pi }%
\left\vert z_{2}\right\rangle \left\langle z_{2}\right\vert e^{d^{\dag
}\left( s_{2}t_{2}-\alpha ^{\ast }r_{2}\right) }\left\vert 0_{d}\right\rangle
\\
&&e^{a^{\dag }\left( \alpha t_{1}+\allowbreak s_{1}r_{1}\right) +b^{\dag
}\left( s_{2}r_{2}+\alpha ^{\ast }\allowbreak t_{2}\right) }\left\vert
0_{a},0_{b}\right\rangle |_{(s_{1},s_{2},s_{3},s_{4})=0}
\end{eqnarray*}%
After a straightforward integration, I finally arrive at the derivative form
of $\left\vert \psi _{LQC}\right\rangle _{ab}$,%
\begin{eqnarray*}
\left\vert \psi _{LQC}\right\rangle _{ab} &=&\frac{1}{\sqrt{p_{cd}}\cosh r}%
\frac{d^{4}}{ds_{1}ds_{2}ds_{3}ds_{4}} \\
&&e^{+\allowbreak s_{3}s_{4}r_{1}r_{2}\tanh
r+s_{1}s_{3}t_{1}+s_{2}s_{4}t_{2}} \\
&&e^{+a^{\dag }\allowbreak s_{1}r_{1}-a^{\dag }s_{4}t_{1}r_{2}\tanh
r+b^{\dag }s_{2}r_{2}-b^{\dag }s_{3}t_{2}r_{1}\tanh r} \\
&&e^{a^{\dag }b^{\dag }t_{1}t_{2}\tanh r}\left\vert 0_{a},0_{b}\right\rangle
|_{(s_{1},s_{2},s_{3},s_{4})=0}.
\end{eqnarray*}%
Therefore the explicit form in Eq.(\ref{1-5}) can be obtained after making
derivation.

\textbf{Appendix B: the density operator }$\rho _{LQC}=\left\vert \psi
_{LQC}\right\rangle _{ab}\left\langle \psi _{LQC}\right\vert $

The conjugate state of $\left\vert \psi _{LQC}\right\rangle _{ab}$\ can be
given by%
\begin{eqnarray*}
&&\left. _{ab}\left\langle \psi _{LQC}\right\vert \right. \\
&=&\frac{1}{\sqrt{p_{cd}}\cosh r}\frac{d^{4}}{dh_{1}dh_{2}dh_{3}dh_{4}}%
\left\langle 0_{a},0_{b}\right\vert e^{abt_{1}t_{2}\tanh r} \\
&&e^{+a\allowbreak h_{1}r_{1}-ah_{4}t_{1}r_{2}\tanh
r+bh_{2}r_{2}-bh_{3}t_{2}r_{1}\tanh r} \\
&&e^{+h_{3}h_{4}r_{1}r_{2}\tanh
r+h_{1}h_{3}t_{1}+h_{2}h_{4}t_{2}}|_{(h_{1},h_{2},h_{3},h_{4})=0}
\end{eqnarray*}%
Then the density operator is%
\begin{eqnarray*}
&&\rho _{LQC} \\
&=&\frac{1}{p_{cd}\cosh ^{2}r}\frac{d^{8}}{%
ds_{1}ds_{2}ds_{3}ds_{4}dh_{1}dh_{2}dh_{3}dh_{4}} \\
&&e^{+\allowbreak s_{3}s_{4}r_{1}r_{2}\tanh
r+s_{1}s_{3}t_{1}+s_{2}s_{4}t_{2}} \\
&&e^{+h_{3}h_{4}r_{1}r_{2}\tanh r+h_{1}h_{3}t_{1}+h_{2}h_{4}t_{2}} \\
&&e^{a^{\dag }\allowbreak s_{1}r_{1}-a^{\dag }s_{4}t_{1}r_{2}\tanh r+b^{\dag
}s_{2}r_{2}-b^{\dag }s_{3}t_{2}r_{1}\tanh r} \\
&&e^{a^{\dag }b^{\dag }t_{1}t_{2}\tanh r}\left\vert 0_{a},0_{b}\right\rangle
\left\langle 0_{a},0_{b}\right\vert e^{abt_{1}t_{2}\tanh r} \\
&&e^{a\allowbreak h_{1}r_{1}-ah_{4}t_{1}r_{2}\tanh
r+bh_{2}r_{2}-bh_{3}t_{2}r_{1}\tanh r} \\
&&|_{(s_{1},s_{2},s_{3},s_{4},h_{1},h_{2},h_{3},h_{4})=0}.
\end{eqnarray*}

\bigskip

\textbf{Appendix C: Success probability of detection}

Due to $\mathrm{Tr}\left( \rho _{LQC}\right) =1$, I have%
\begin{eqnarray*}
p_{cd} &=&\frac{\cosh ^{2}\lambda }{\cosh ^{2}r}\frac{d^{8}}{%
ds_{1}ds_{2}ds_{3}ds_{4}dh_{1}dh_{2}dh_{3}dh_{4}} \\
&&\exp \left( \Xi \right)
|_{(s_{1},s_{2},s_{3},s_{4},h_{1},h_{2},h_{3},h_{4})=0},
\end{eqnarray*}%
where I have set%
\begin{eqnarray*}
\Xi &=&+\epsilon _{1}\allowbreak \left( s_{3}s_{4}+h_{3}h_{4}\right)
+\epsilon _{2}\left( s_{1}s_{2}+\allowbreak h_{1}h_{2}\right) \\
&&+\epsilon _{3}\left( s_{1}s_{3}+h_{1}h_{3}\right) +\epsilon _{4}\left(
s_{2}s_{4}+h_{2}h_{4}\right) \\
&&-\epsilon _{5}\left( \allowbreak h_{2}s_{3}+s_{2}\allowbreak h_{3}\right)
-\epsilon _{6}\left( s_{4}h_{1}\allowbreak +s_{1}h_{4}\right) \\
&&+\epsilon _{7}s_{1}\allowbreak h_{1}+\epsilon _{8}s_{2}h_{2}+\epsilon
_{9}s_{3}h_{3}+\epsilon _{10}s_{4}h_{4}
\end{eqnarray*}%
with$\allowbreak $%
\begin{eqnarray*}
\epsilon _{1} &=&\frac{r_{1}r_{2}\sinh 2\lambda }{2t_{1}t_{2}},\epsilon _{2}=%
\frac{r_{1}r_{2}\sinh 2\lambda }{2}, \\
\epsilon _{3} &=&t_{1}\cosh ^{2}\lambda -\frac{\sinh ^{2}\lambda }{t_{1}}, \\
\epsilon _{4} &=&t_{2}\cosh ^{2}\lambda -\frac{\sinh ^{2}\lambda }{t_{2}} \\
\epsilon _{5} &=&\frac{r_{1}r_{2}\sinh 2\lambda }{2t_{1}},\epsilon _{6}=%
\frac{r_{1}r_{2}\sinh 2\lambda }{2t_{2}} \\
\epsilon _{7} &=&r_{1}^{2}\cosh ^{2}\lambda ,\epsilon _{8}=r_{2}^{2}\cosh
^{2}\lambda \\
\epsilon _{9} &=&\frac{r_{1}^{2}\sinh ^{2}\lambda }{t_{1}^{2}},\epsilon
_{10}=\frac{r_{2}^{2}\sinh ^{2}\lambda }{t_{2}^{2}}.
\end{eqnarray*}

\textbf{Appendix D: Expectation value }$\left\langle a^{\dag
k_{1}}\allowbreak b^{\dag k_{2}}a^{l_{1}}\allowbreak b^{l_{2}}\right\rangle $

According to $\left\langle a^{\dag k_{1}}\allowbreak b^{\dag
k_{2}}a^{l_{1}}\allowbreak b^{l_{2}}\right\rangle =\mathrm{Tr}\left( a^{\dag
k_{1}}\allowbreak b^{\dag k_{2}}a^{l_{1}}\allowbreak b^{l_{2}}\rho
_{LQC}\right) $ and making detailed calculation, I obtain%
\begin{eqnarray*}
&&\left\langle a^{\dag k_{1}}\allowbreak b^{\dag k_{2}}a^{l_{1}}\allowbreak
b^{l_{2}}\right\rangle \\
&=&\frac{\cosh ^{2}\lambda }{p_{cd}\cosh ^{2}r}\frac{%
d^{8+k_{1}+l_{1}+k_{2}+l_{2}}}{%
ds_{1}ds_{2}ds_{3}ds_{4}dh_{1}dh_{2}dh_{3}dh_{4}df_{1}^{k_{1}}df_{2}^{l_{1}}dg_{1}^{k_{2}}dg_{2}^{l_{2}}%
} \\
&&\exp \left( \Xi +\Theta \right)
|_{(s_{1},s_{2},s_{3},s_{4},h_{1},h_{2},h_{3},h_{4},f_{1},f_{2},g_{1},g_{2})=0,}
\end{eqnarray*}%
where I have set%
\begin{eqnarray*}
\Theta &=&+\eta _{1}\left( s_{1}g_{1}+h_{1}g_{2}\right) +\eta _{2}\left(
s_{2}f_{1}+h_{2}f_{2}\right) \\
&&+\eta _{3}\left( s_{2}g_{2}+h_{2}g_{1}\right) +\eta _{4}\left( \allowbreak
s_{1}f_{2}+h_{1}f_{1}\right) \\
&&-\eta _{5}\left( s_{3}g_{2}+h_{3}g_{1}\right) -\eta _{6}\left(
s_{4}f_{2}+\allowbreak h_{4}f_{1}\right) \\
&&-\eta _{7}\left( s_{4}g_{1}+h_{4}g_{2}\right) -\eta _{8}\left( \allowbreak
s_{3}f_{1}+h_{3}f_{2}\right) \\
&&+\eta _{9}\left( f_{1}g_{1}+f_{2}g_{2}\right) +\eta _{10}\left(
f_{1}f_{2}+\allowbreak g_{1}g_{2}\right)
\end{eqnarray*}%
with$\allowbreak $%
\begin{eqnarray*}
\eta _{1} &=&\frac{r_{1}\sinh 2\lambda }{2},\eta _{2}=\frac{r_{2}\sinh
2\lambda }{2}, \\
\eta _{3} &=&r_{2}\cosh ^{2}\lambda ,\eta _{4}=r_{1}\cosh ^{2}\lambda , \\
\eta _{5} &=&\frac{r_{1}\sinh 2\lambda }{2t_{1}}\allowbreak ,\eta _{6}=\frac{%
r_{2}\sinh 2\lambda }{2t_{2}}, \\
\eta _{7} &=&\frac{r_{2}\sinh ^{2}\lambda }{t_{2}},\eta _{8}=\frac{%
r_{1}\sinh ^{2}\lambda }{t_{1}}, \\
\text{ }\eta _{9} &=&\allowbreak \frac{\sinh 2\lambda }{2},\eta _{10}=\sinh
^{2}\lambda .
\end{eqnarray*}

\textbf{Appendix E:. the expressions of }$x_{i}$\textbf{, }$y_{i}$\textbf{, }%
$z_{i}$

Here I list the expressions of $x_{i}$, $y_{i}$, $z_{i}$ as follow 
\begin{eqnarray*}
x_{0} &=&-2t_{1}^{2}t_{2}^{4}+2t_{1}^{4}t_{2}^{4}, \\
x_{1}
&=&1-4t_{1}^{2}+4t_{1}^{4}-4t_{2}^{2}+4t_{2}^{4}+16t_{1}^{2}t_{2}^{2}-16t_{1}^{4}t_{2}^{2}
\\
&&-14t_{1}^{2}t_{2}^{4}+14t_{1}^{4}t_{2}^{4}+t_{1}^{2}t_{2}^{6}-2t_{1}^{4}t_{2}^{6}+t_{1}^{6}t_{2}^{6},
\\
x_{2}
&=&4t_{1}^{2}t_{2}^{2}-12t_{1}^{4}t_{2}^{2}+8t_{1}^{6}t_{2}^{2}-16t_{1}^{2}t_{2}^{4}+48t_{1}^{4}t_{2}^{4}
\\
&&-32t_{1}^{6}t_{2}^{4}+14t_{1}^{2}t_{2}^{6}-34t_{1}^{4}t_{2}^{6}+20t_{1}^{6}t_{2}^{6},
\\
x_{3}
&=&22t_{1}^{2}t_{2}^{2}-60t_{1}^{4}t_{2}^{2}+40t_{1}^{6}t_{2}^{2}-56t_{1}^{2}t_{2}^{4}+146t_{1}^{4}t_{2}^{4}
\\
&&-92t_{1}^{6}t_{2}^{4}+2t_{1}^{8}t_{2}^{4}+33t_{1}^{2}t_{2}^{6}-92t_{1}^{4}t_{2}^{6}+61t_{1}^{6}t_{2}^{6}
\\
&&-8t_{1}^{8}t_{2}^{6}+4t_{1}^{4}t_{2}^{8}-8t_{1}^{6}t_{2}^{8}+4t_{1}^{8}t_{2}^{8},
\\
x_{4}
&=&24t_{1}^{4}t_{2}^{4}-52t_{1}^{6}t_{2}^{4}+28t_{1}^{8}t_{2}^{4}-48t_{1}^{4}t_{2}^{6}+88t_{1}^{6}t_{2}^{6}
\\
&&-40t_{1}^{8}t_{2}^{6}+20t_{1}^{4}t_{2}^{8}-34t_{1}^{6}t_{2}^{8}+14t_{1}^{8}t_{2}^{8},
\\
x_{5}
&=&40t_{1}^{4}t_{2}^{4}-76t_{1}^{6}t_{2}^{4}+30t_{1}^{8}t_{2}^{4}-76t_{1}^{4}t_{2}^{6}+140t_{1}^{6}t_{2}^{6}
\\
&&-56t_{1}^{8}t_{2}^{6}+4t_{1}^{10}t_{2}^{6}+36t_{1}^{4}t_{2}^{8}-58t_{1}^{6}t_{2}^{8}+26t_{1}^{8}t_{2}^{8}
\\
&&-4t_{1}^{10}t_{2}^{8}+t_{1}^{6}t_{2}^{10}-2t_{1}^{8}t_{2}^{10}+t_{1}^{10}t_{2}^{10},
\\
x_{6}
&=&8t_{1}^{6}t_{2}^{6}-8t_{1}^{8}t_{2}^{6}-8t_{1}^{6}t_{2}^{8}+8t_{1}^{8}t_{2}^{8}+2t_{1}^{6}t_{2}^{10}-2t_{1}^{8}t_{2}^{10},
\\
x_{7}
&=&14t_{1}^{6}t_{2}^{6}-16t_{1}^{8}t_{2}^{6}+4t_{1}^{10}t_{2}^{6}-20t_{1}^{6}t_{2}^{8}+16t_{1}^{8}t_{2}^{8}
\\
&&-4t_{1}^{10}t_{2}^{8}+5t_{1}^{6}t_{2}^{10}-4t_{1}^{8}t_{2}^{10}+t_{1}^{10}t_{2}^{10},
\\
x_{8} &=&0,\text{ \ \ }x_{9}=t_{1}^{8}t_{2}^{8},
\end{eqnarray*}%
\begin{eqnarray*}
y_{0} &=&-2t_{1}^{4}t_{2}^{2}+2t_{1}^{4}t_{2}^{4}, \\
y_{1}
&=&1-4t_{1}^{2}+4t_{1}^{4}-4t_{2}^{2}+4t_{2}^{4}+16t_{1}^{2}t_{2}^{2}-16t_{1}^{2}t_{2}^{4}
\\
&&-14t_{1}^{4}t_{2}^{2}+14t_{1}^{4}t_{2}^{4}+t_{1}^{6}t_{2}^{2}-2t_{1}^{6}t_{2}^{4}+t_{1}^{6}t_{2}^{6},
\\
y_{2}
&=&4t_{1}^{2}t_{2}^{2}-16t_{1}^{4}t_{2}^{2}+14t_{1}^{6}t_{2}^{2}-12t_{1}^{2}t_{2}^{4}+48t_{1}^{4}t_{2}^{4}
\\
&&-34t_{1}^{6}t_{2}^{4}+8t_{1}^{2}t_{2}^{6}-32t_{1}^{4}t_{2}^{6}+20t_{1}^{6}t_{2}^{6},
\\
y_{3}
&=&22t_{1}^{2}t_{2}^{2}-56t_{1}^{4}t_{2}^{2}+33t_{1}^{6}t_{2}^{2}-60t_{1}^{2}t_{2}^{4}+146t_{1}^{4}t_{2}^{4}
\\
&&-92t_{1}^{6}t_{2}^{4}+4t_{1}^{8}t_{2}^{4}+40t_{1}^{2}t_{2}^{6}-92t_{1}^{4}t_{2}^{6}+61t_{1}^{6}t_{2}^{6}
\\
&&-8t_{1}^{8}t_{2}^{6}+2t_{1}^{4}t_{2}^{8}-8t_{1}^{6}t_{2}^{8}+4t_{1}^{8}t_{2}^{8},
\\
y_{4}
&=&24t_{1}^{4}t_{2}^{4}-48t_{1}^{6}t_{2}^{4}+20t_{1}^{8}t_{2}^{4}-52t_{1}^{4}t_{2}^{6}+88t_{1}^{6}t_{2}^{6}
\\
&&-34t_{1}^{8}t_{2}^{6}+28t_{1}^{4}t_{2}^{8}-40t_{1}^{6}t_{2}^{8}+14t_{1}^{8}t_{2}^{8},
\\
y_{5}
&=&40t_{1}^{4}t_{2}^{4}-76t_{1}^{6}t_{2}^{4}+36t_{1}^{8}t_{2}^{4}-76t_{1}^{4}t_{2}^{6}+140t_{1}^{6}t_{2}^{6}
\\
&&-58t_{1}^{8}t_{2}^{6}+t_{1}^{10}t_{2}^{6}+30t_{1}^{4}t_{2}^{8}-56t_{1}^{6}t_{2}^{8}+26t_{1}^{8}t_{2}^{8}
\\
&&-2t_{1}^{10}t_{2}^{8}+4t_{1}^{6}t_{2}^{10}-4t_{1}^{8}t_{2}^{10}+t_{1}^{10}t_{2}^{10},
\\
y_{6}
&=&8t_{1}^{6}t_{2}^{6}-8t_{1}^{8}t_{2}^{6}-8t_{1}^{6}t_{2}^{8}+8t_{1}^{8}t_{2}^{8}+2t_{1}^{10}t_{2}^{6}-2t_{1}^{10}t_{2}^{8},
\\
y_{7}
&=&14t_{1}^{6}t_{2}^{6}-20t_{1}^{8}t_{2}^{6}+5t_{1}^{10}t_{2}^{6}-16t_{1}^{6}t_{2}^{8}+16t_{1}^{8}t_{2}^{8}
\\
&&-4t_{1}^{10}t_{2}^{8}+4t_{1}^{6}t_{2}^{10}-4t_{1}^{8}t_{2}^{10}+t_{1}^{10}t_{2}^{10},
\\
y_{8} &=&0,\text{ \ \ }y_{9}=t_{1}^{8}t_{2}^{8},
\end{eqnarray*}%
as well as%
\begin{eqnarray*}
z_{0} &=&1-2t_{1}^{2}-2t_{2}^{2}+4t_{1}^{2}t_{2}^{2}, \\
z_{1}
&=&-t_{1}^{2}+2t_{1}^{4}-t_{2}^{2}+2t_{2}^{4}+6t_{1}^{2}t_{2}^{2}-8t_{1}^{4}t_{2}^{2}
\\
&&-8t_{1}^{2}t_{2}^{4}+8t_{1}^{4}t_{2}^{4}, \\
z_{2}
&=&8-27t_{1}^{2}+22t_{1}^{4}-27t_{2}^{2}+22t_{2}^{4}+87t_{1}^{2}t_{2}^{2} \\
&&-67t_{1}^{4}t_{2}^{2}+2t_{1}^{6}t_{2}^{2}-67t_{1}^{2}t_{2}^{4}+49t_{1}^{4}t_{2}^{4}-6t_{1}^{6}t_{2}^{4}
\\
&&+2t_{1}^{2}t_{2}^{6}-6t_{1}^{4}t_{2}^{6}+4t_{1}^{6}t_{2}^{6}, \\
z_{3}
&=&14t_{1}^{2}t_{2}^{2}-37t_{1}^{4}t_{2}^{2}+22t_{1}^{6}t_{2}^{2}-37t_{1}^{2}t_{2}^{4}+92t_{1}^{4}t_{2}^{4}
\\
&&-50t_{1}^{6}t_{2}^{4}+22t_{1}^{2}t_{2}^{6}-50t_{1}^{4}t_{2}^{6}+24t_{1}^{6}t_{2}^{6},
\\
z_{4}
&=&45t_{1}^{2}t_{2}^{2}-98t_{1}^{4}t_{2}^{2}+48t_{1}^{6}t_{2}^{2}-98t_{1}^{2}t_{2}^{4}+197t_{1}^{4}t_{2}^{4}
\\
&&-90t_{1}^{6}t_{2}^{4}+4t_{1}^{8}t_{2}^{4}+48t_{1}^{2}t_{2}^{6}-90t_{1}^{4}t_{2}^{6}+46t_{1}^{6}t_{2}^{6}
\\
&&-6t_{1}^{8}t_{2}^{6}+4t_{1}^{4}t_{2}^{8}-6t_{1}^{6}t_{2}^{8}+2t_{1}^{8}t_{2}^{8},
\\
z_{5}
&=&20t_{1}^{4}t_{2}^{4}-33t_{1}^{6}t_{2}^{4}+12t_{1}^{8}t_{2}^{4}-33t_{1}^{4}t_{2}^{6}+46t_{1}^{6}t_{2}^{6}
\\
&&-14t_{1}^{8}t_{2}^{6}+12t_{1}^{4}t_{2}^{8}-14t_{1}^{6}t_{2}^{8}+4t_{1}^{8}t_{2}^{8},
\\
z_{6}
&=&24t_{1}^{4}t_{2}^{4}-31t_{1}^{6}t_{2}^{4}+8t_{1}^{8}t_{2}^{4}-31t_{1}^{4}t_{2}^{6}+33t_{1}^{6}t_{2}^{6}
\\
&&-9t_{1}^{8}t_{2}^{6}+8t_{1}^{4}t_{2}^{8}-9t_{1}^{6}t_{2}^{8}+3t_{1}^{8}t_{2}^{8},
\\
z_{7} &=&2t_{1}^{6}t_{2}^{6}-t_{1}^{8}t_{2}^{6}-t_{1}^{6}t_{2}^{8},\text{ \
\ }z_{8}=t_{1}^{6}t_{2}^{6}.
\end{eqnarray*}

\textbf{Appendix F: Characteristic function of }$\left\vert \psi
_{LQC}\right\rangle _{ab}$

Noticing the displacement operators $D_{a}\left( \alpha \right) =e^{\frac{%
\left\vert \alpha \right\vert ^{2}}{2}}e^{-\alpha ^{\ast }a}e^{\alpha
a^{\dag }},D_{b}\left( \beta \right) =e^{\frac{\left\vert \beta \right\vert
^{2}}{2}}e^{-\beta ^{\ast }b}e^{\beta b^{\dag }}$, the CF of $\left\vert
\psi _{LQC}\right\rangle _{ab}$\ can be calculated as 
\begin{eqnarray*}
&&\chi _{E}\left( \alpha ,\beta \right) \\
&=&\frac{\cosh ^{2}\lambda }{p_{cd}\cosh ^{2}r}\frac{d^{8}}{%
ds_{1}ds_{2}ds_{3}ds_{4}dh_{1}dh_{2}dh_{3}dh_{4}} \\
&&e^{\Xi -\Lambda \left\vert \alpha \right\vert ^{2}+\chi _{\alpha }\alpha
+\chi _{\alpha ^{\ast }}\allowbreak \alpha ^{\ast }-\Lambda \left\vert \beta
\right\vert ^{2}+\chi _{\beta }\beta +\chi _{\beta ^{\ast }}\beta ^{\ast
}+\eta _{9}\left( \alpha \beta +\alpha ^{\ast }\beta ^{\ast }\right) } \\
&&|_{(s_{1},s_{2},s_{3},s_{4},h_{1},h_{2},h_{3},h_{4})=0},
\end{eqnarray*}%
where I have set $\Lambda =\cosh ^{2}\lambda -\frac{1}{2}$ and%
\begin{eqnarray*}
\chi _{\alpha } &=&h_{1}\eta _{4}\allowbreak +s_{2}\eta _{2}\allowbreak
-s_{3}\allowbreak \eta _{8}-h_{4}\eta _{6}, \\
\chi _{\alpha ^{\ast }} &=&-s_{1}\eta _{4}-h_{2}\eta _{2}+h_{3}\eta
_{8}+s_{4}\eta _{6}, \\
\chi _{\beta } &=&s_{1}\eta _{1}+h_{2}\eta _{3}-h_{3}\eta _{5}-s_{4}\eta
_{7}, \\
\chi _{\beta ^{\ast }} &=&-\allowbreak h_{1}\eta _{1}-s_{2}\eta
_{3}+s_{3}\allowbreak \eta _{5}+\allowbreak h_{4}\eta _{7}.
\end{eqnarray*}

\textbf{Appendix G: the fidelity of QT of CVs}

Considering the entangled sate $\left\vert \psi _{LQC}\right\rangle _{ab}$
to teleport a coherent (vacuum) state and substituting $\chi _{in}(z)=\exp
[-|z|^{2}/2]$ and $\chi _{out}(z)=\chi _{in}(z)\chi _{E}\left( z^{\ast
},z\right) $ into $F=\int \frac{d^{2}z}{\pi }\chi _{in}(-z)\chi _{out}(z)$
yields%
\begin{eqnarray*}
F &=&\frac{\kappa _{0}}{p_{cd}\cosh ^{2}r}\frac{d^{8}}{%
ds_{1}ds_{2}ds_{3}ds_{4}dh_{1}dh_{2}dh_{3}dh_{4}} \\
&&\exp \left( \Pi \right)
|_{(s_{1},s_{2},s_{3},s_{4},h_{1},h_{2},h_{3},h_{4})=0},
\end{eqnarray*}%
where I have set%
\begin{eqnarray*}
\Pi  &=&+\kappa _{1}\left( s_{1}s_{2}+h_{1}h_{2}\right) +\kappa _{2}\left(
\allowbreak s_{3}s_{4}+h_{3}h_{4}\right)  \\
&&+\kappa _{3}\left( s_{1}s_{3}+h_{1}h_{3}\right) +\kappa _{4}\left(
s_{2}s_{4}+h_{2}h_{4}\right)  \\
&&-\kappa _{5}\left( s_{3}h_{2}+s_{2}\allowbreak h_{3}\right) -\kappa
_{6}\left( s_{4}h_{1}+s_{1}h_{4}\right)  \\
&&+\kappa _{7}s_{1}h_{1}+\kappa _{8}s_{2}\allowbreak h_{2}+\kappa
_{9}s_{3}h_{3}+\kappa _{10}s_{4}h_{4}.
\end{eqnarray*}%
with $\kappa _{0}=[2\left( 1-\tanh \lambda \right) ]^{-1}$ and%
\begin{eqnarray*}
\kappa _{1} &=&\kappa _{0}r_{1}\allowbreak r_{2},\kappa _{2}=(\kappa _{0}+%
\frac{1}{2})r_{1}r_{2}\tanh r, \\
\kappa _{3} &=&\frac{\allowbreak t_{1}\sinh 2\lambda }{8\kappa _{0}}-\frac{%
\kappa _{0}\tanh \lambda }{t_{1}}+t_{1}\cosh ^{2}\lambda , \\
\kappa _{4} &=&\frac{\allowbreak t_{2}\sinh 2\lambda }{8\kappa _{0}}-\frac{%
\kappa _{0}\tanh \lambda }{t_{2}}+t_{2}\cosh ^{2}\lambda  \\
\kappa _{5} &=&\frac{\kappa _{0}r_{1}r_{2}\tanh \lambda }{t_{1}},\kappa _{6}=%
\frac{\kappa _{0}r_{1}r_{2}\tanh \lambda }{t_{2}}, \\
\kappa _{7} &=&\kappa _{0}r_{1}^{2},\kappa _{8}=\kappa _{0}r_{2}^{2} \\
\kappa _{9} &=&\frac{\kappa _{0}r_{1}^{2}\tanh ^{2}\lambda }{t_{1}^{2}}%
,\kappa _{10}=\frac{\kappa _{0}r_{2}^{2}\tanh ^{2}\lambda }{t_{2}^{2}}
\end{eqnarray*}%
Thus Eq.(\ref{3-3}) can be obtained.$\allowbreak $

\end{document}